\documentclass[a4paper]{article}

\usepackage[backend=biber,style=numeric,  
            giveninits=true,
            doi=false,isbn=false,url=false,
            eprint=false,maxbibnames=99]{biblatex}
\usepackage[english]{babel}
\usepackage[utf8]{inputenc}
\usepackage[colorinlistoftodos]{todonotes}
\usepackage{amsmath}
\usepackage{amssymb}
\usepackage{amsthm}
\usepackage{mathtools}
\usepackage{mathrsfs}
\usepackage{siunitx}
\usepackage{algpseudocode}

\usepackage{tikz}

\usepackage{tabularx}
\usepackage{booktabs}
\usepackage{multirow}

\usepackage{empheq}

\usepackage{lipsum}

\usepackage[toc,page]{appendix}

\usepackage{rotating}

\usepackage{makecell}

\usepackage{lscape}
\usepackage{xcolor,colortbl}
\definecolor{lavender}{rgb}{0.9, 0.9, 0.98}
\everymath{\displaystyle}
\usepackage{float}
\usepackage[font={small}]{caption}
\usepackage[a4paper, total={6in, 8in}]{geometry}
\usepackage{csquotes}
\usepackage{subcaption}

\newif\ifdouble

\newcommand{\papertitle}{Full-field surrogate modeling of cardiac function encoding geometric variability}

\newcommand{\keywordOne}{Computational Cardiology}
\newcommand{\keywordTwo}{Scientific Machine Learning}
\newcommand{\keywordThree}{Statistical Shape Modeling}

\newcommand{\keywordFive}{Tetralogy of Fallot}
\newcommand{\keywordSix}{Cardiac Electrophysiology}
\newcommand{\keywordSeven}{Surrogate Modeling}

\newcommand{\param}{\boldsymbol{\theta}}

\newcommand{\paramBiV}{\param_\mathrm{BiV}}

\newcommand{\paramBiVAdim}{\widetilde{\param}_\mathrm{BiV}}

\newcommand{\fZero}{{\boldsymbol{f}_0}}

\newcommand{\Iion}{{\mathcal{I}_{\mathrm{ion}}}}
\newcommand{\Iapp}{{\mathcal{I}_{\mathrm{app}}}}
\newcommand{\DiffTens}{\boldsymbol{D}_{\mathrm{M}}}
\newcommand{\Pot}{u}

\newcommand{\Ionic}{\boldsymbol{y}}

\newcommand{\RhsIonic}{\boldsymbol{F}}

\newcommand{\Da}{D_\mathrm{ani}}
\newcommand{\Di}{D_\mathrm{iso}}
\newcommand{\Dp}{D_\mathrm{purk}}

\newcommand{\IdentityVec}{\mathbf{I}}

\newcommand{\NumPatients} {N_\mathcal{O}}
\newcommand{\syntheticPatients} {N_\mathcal{S}}
\newcommand{\totalPatients} {N_\mathcal{T}}

\newcommand{\paramSpace}{\boldsymbol{\Theta}}

\newcommand{\ANNState}{\mathbf{z}}

\newcommand{\ANNStateTildeObs}{\widetilde{z}_\mathrm{obs}}

\newcommand{\ANNLatent}{\mathbf{z}_\mathrm{latent}}

\newcommand{\ANNparam}{\mathbf{w}}
\newcommand{\ANNparamTrained}{\widehat{\ANNparam}}

\newcommand{\ANNStateAT}{z_\mathrm{AT}}
\newcommand{\ANNRhs}{\mathcal{B L \kern0.05em N \kern-0.05em M}}
\newcommand{\NN}{\mathcal{N \kern-0.05em N}}

\newcommand{\ANNStateATAdim}{\widetilde{z}_\mathrm{AT}}

\newcommand{\NumANNState}{N_\mathrm{z}}

\newcommand{\NumANNWeights}{N_\mathrm{w}}

\newcommand{\NumModes}{N_\mathcal{M}}

\graphicspath{{pictures/}}
\doublefalse

\title{{\papertitle}}
\author{Elena Sabdy Martinez$^{1, 2, *}$, Beatrice Moscoloni$^{4,5}$, Matteo Salvador$^{1, 2, 3, 6}$, \\ Fanwei Kong$^{1, 2, 3}$, Mathias Peirlinck$^{4}$, Alison Lesley Marsden$^{1, 2, 3, 7}$}

\date{\footnotesize
    $^1$ Institute for Computational and Mathematical Engineering, Stanford University, CA, USA \\
    $^2$ Cardiovascular Institute, Stanford University, CA, USA \\
    $^3$ Pediatric Cardiology, Stanford University, CA, USA \\
    $^4$ Department of BioMechanical Engineering, Delft University of Technology, Delft, Netherlands \\
    $^6$ IBiTech-bioMMeda, Ghent University, Ghent, Belgium \\
    $^6$ Pasteur Labs, Brooklyn, NY 11205, USA \\
    $^7$ Department of Bioengineering, Stanford University, CA, USA \\
    $^*$ \textit{Corresponding author} (\texttt{elenasm@stanford.edu}) \\
}

\begin{document}
	\maketitle
	
	\begin{abstract}
		Combining physics-based modeling with data-driven methods is critical to enabling the translation of computational methods to clinical use in cardiology. The use of rigorous differential equations combined with machine learning tools allows for model personalization with uncertainty quantification in time frames compatible with clinical practice. However, accurate and efficient surrogate models of cardiac function, built from physics-based numerical simulation, are still mostly geometry-specific and require retraining for different patients and pathological conditions. We propose a novel computational pipeline to embed cardiac anatomies into full-field surrogate models. We generate a dataset of electrophysiology simulations using a complex multi-scale mathematical model coupling partial and ordinary differential equations. We adopt Branched Latent Neural Maps (BLNMs) as an effective scientific machine learning method to encode activation maps extracted from physics-based numerical simulations into a neural network. Leveraging large deformation diffeomorphic metric mappings, we build a biventricular anatomical atlas and parametrize the anatomical variability of a small and challenging cohort of 13 pediatric patients affected by Tetralogy of Fallot. We propose a novel statistical shape modeling based z-score sampling approach to generate a new synthetic cohort of 52 biventricular geometries that are compatible with the original geometrical variability. This synthetic cohort acts as the training set for BLNMs. Our surrogate model demonstrates robustness and great generalization across the complex original patient cohort, achieving an average adimensional mean squared error of 0.0034. The Python implementation of our BLNM model is publicly available under MIT License at \url{https://github.com/StanfordCBCL/BLNM}.
	\end{abstract}
	
        \newpage
	\section{Introduction}
\label{sec:introduction}
For decades, cardiologists have based treatment decisions on a ``one-size-fits-all'' assumption that patients with standardized symptoms must share the same disease pathophenotype and thus exhibit similar responses to the conventional treatment plan \cite{Leopold}. However, the significant variation among heart physiology, geometry, and cell-to-organ scale properties often leads to diverse responses to the same interventions, particularly in pediatric cardiology \cite{Tikenogullari2023}. For example, a single action potential model derived from a dataset cannot adequately capture the variability observed in experimentally measured action potentials, even within a scientifically homogeneous population \cite{Britton2013}. This limitation becomes particularly crucial in clinical practice. Treating cardiac arrhythmias using a general model rather than an individualized approach can result in incorrect defibrillation energy settings or medication doses \cite{trayanova2013,Peirlinck2022}. This underscores the critical need for personalized models to improve patient safety and treatment efficacy \cite{Vadakklumpadan2010,Mirams2020}.

Physics-based mathematical models can incorporate both patient-specific anatomic imaging and clinical data. These models employ rigorous differential equations to simulate cardiac dynamics while accounting for individual variations in heart structure \cite{Peirlinck2021}. Integrating data-driven methods with physics-based models enables the inclusion of diverse datasets obtained from various clinical modalities, such as imaging (e.g MRI or CT scans), electrophysiology recordings (e.g. ECG or EEG), and hemodynamic measurements (e.g blood flow or pressure data). This accelerates parameter calibration and facilitates more informed and timely decision-making in clinical settings \cite{Niederer2019, Kadem2023, Jaffery2024}. Recent works demonstrate the efficacy of applying machine learning techniques to multi-scale cardiac models tailored to individual patients \cite{GiffardRoisin2017, Doste2019, Manzoni2020,Peirlinck2021a,Roney2022,Regazzoni2022_2, Cicci2023, Pegolotti2024}. Various methods have aimed to reduce the computational cost of physics-based simulations while achieving accurate and efficient model performance specifically for electrophysiology \cite{Salvador2024}. For example, reduced-order models (ROMs) based on artificial neural networks (ANNs) can accurately generate active force in patient-specific geometries while maintaining computational efficiency \cite{Regazzoni2020}. Deep learning-based reduced order models (DL-ROMs) for cardiac electrophysiology can efficiently evaluate outputs of clinical interest such as activation maps and action potentials \cite{Fresca2020}. By incorporating the PDEs associated with activation mapping and neural networks, Physics-Informed Neural Networks (PINNs) can create activation maps based on a few sparse activation measurements \cite{Costabal2020}. These models, however, often overlook the significant interpatient variability in cardiac geometries, a crucial factor for accurate scalable predictions \cite{Cherry2011}.

Statistical shape modeling (SSM) has been extensively applied to characterize anatomical variability across various cardiovascular applications \cite{rodero_2021_4chamber_ssm, nagel_2021_biatrial_ssm, kollar_2022_rv_ssm, sophocleus_2022_implications_lv, cutugno_2021_lv_remodeling,Moscoloni2025}. 
Typically, SSM construction involves  first mapping a cohort of anatomical meshes onto a common template to ensure vertex-to-vertex correspondence among subjects. Subsequently, dimensionality reduction techniques - most commonly principal component analysis (PCA) - are applied to derive a compact, low-dimensional representation of the cohort. PCA identifies an average anatomical configuration and  principal components (or modes) that capture the majority of the anatomical variability present within the dataset.  Following PCA, each anatomical mesh can be reconstructed as a linear combination of these modes, enabling the encoding of individual geomtrical varibility through mode-specific weights. 
Recently, several studies have combined statistical shape modeling with advanced sampling strategies to generate large synthetic cohorts of cardiac surface meshes \cite{romero_2021_synthetic_gen, thamsen_2021_synthetic_gen, bridio_2023_gen_virtual_AIS, ostendorf_2024_synthetic_gen, saitta_2023_paperludo, scuoppo_2024_virtual_cohort}. 
The generation of synthetic anatomical cohorts holds significant potential for in silico studies and surrogate model training. However, realizing this potential critically depends on well-designed sampling strategies and rigorous acceptance criteria to ensure that the generated synthetic cohorts excludes anatomies that are physiologically implausible or improbable \cite{romero_2021_synthetic_gen, niederer_2020_heart_virtual_cohort}. 

This paper proposes a novel computational pipeline for predicting activation maps from electrophysiology simulations by embedding cardiac anatomies into full-field surrogate models using Branched Latent Neural Maps (BLNMs). BLNMs use feedforward partially-connected neural networks to separate contributions from distinct inputs, thereby maintaining training/inference computational efficiency and scalable memory footprints while accurately learning complex input-output relationships with significant variability \cite{Salvador2024BLNM}. Our approach uniquely integrates innovative SSM approaches with BLNMs to capture the geometric and functional variability of highly variant cardiac geometries. 

We consider a cohort of 13 pediatric patients affected by Tetralogy of Fallot. To expand our dataset, we generate a synthetic cohort of 52 additional biventricular geometries that align with the original distribution. For both the original and synthetic cohort, we produce a dataset of electrophysiology simulations using a multi-scale mathematical model based on differential equations. Using BLNMs, we encode the activation maps extracted from these physics-based numerical simulations. These activation maps represent the timing of electrical activation in cardiac tissue and are key in guiding medical treatments and procedures for various cardiac conditions \cite{Karoui2021}. By testing this framework on a challenging cohort of patients, we demonstrate its robustness in handling diverse anatomies while preserving computational efficiency and without needing model modification.
	\section{Methods}
\label{sec:methods}
\begin{figure}[H]
  \centering
  \includegraphics[width=1.0\textwidth]{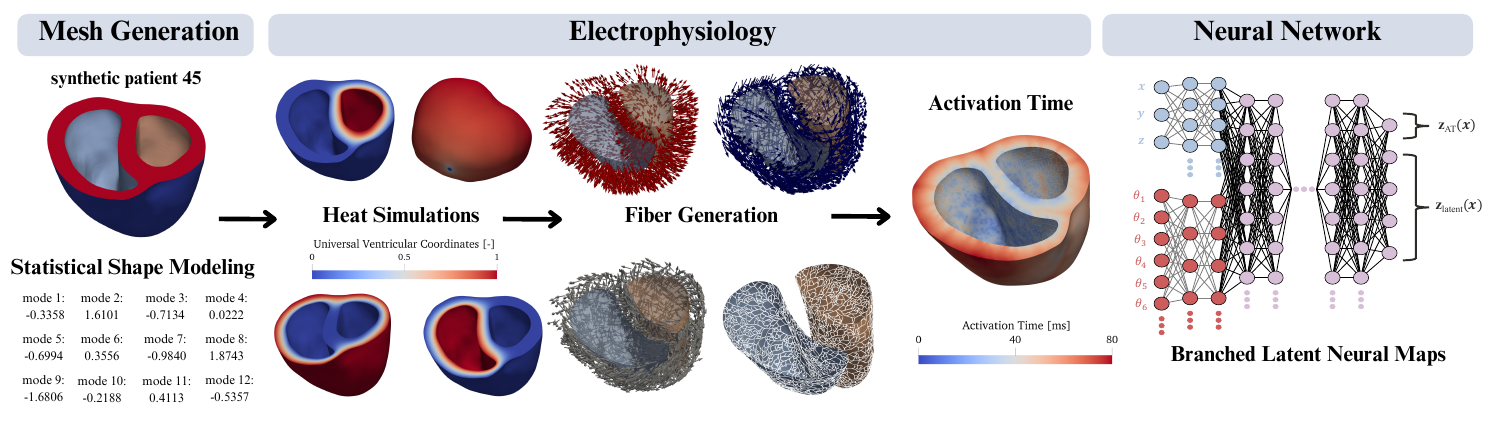}
   \caption{Sketch of the computational pipeline illustrated for synthetic patient 45. We use SSM to generate a synthetic patient by sampling 12 modes. For this patient, we conduct heat simulations to define some universal ventricular coordinates \cite{Bayer2012}. Next, we create the corresponding fiber fields: fiber sheets (top left), longitudinal fibers (top right), normal fibers (bottom left), and Purkinje fibers (bottom right). We use these fiber orientations to run an electrophysiology simulation and generate the activation map.  BLNMs predict activation maps by using two branches for modes and spatial coordinates, respectively}
\end{figure}

We introduce a methodological framework for predicting cardiac
activation maps in highly heterogeneous biventricular anatomies by simulating cardiac electrophysiology driven by anatomical form. Although this framework is applicable to a wide range of healthy and diseased heart morphologies, we specifically highlight its application to a challenging cohort of Tetralogy of Fallot (ToF) patients. ToF is a complex congenital heart disease (CHD) characterized by four key anatomical abnormalities: a ventricular septal defect, an overriding aorta, stenosis of the right ventricular outflow tract, and right ventricular hypertrophy. Given the complex nature and varying severity of these defects, ToF presents an ideal pathophysiology to showcase the capabilities of our computational pipeline \cite{Khan2018}. This process can be divided into three parts: first, generating a cohort of 52 synthetic geometries derived from 13 ToF patients; second, creating geometry-specific cardiac activation maps; and third, using Branched Latent Neural Maps to predict activation maps for different geometries.

\subsection{Original Cohort}
We utilize ImageCHD, a dataset consisting of 110 3D computed tomography (CT) images that capture of a wide range of congenital heart diseases in patients between 1 month and 40 years of age \cite{chd_database}. For this study, we focus specifically on the subset of patients with Tetralogy of Fallot (ToF). We generate high-resolution anatomical meshes from the ground truth myocardium segmentations of these 3D CT images. We focus on modeling only the ventricular region below the ventricular septal defect (VSD) in these patients to concentrate on areas most impacted by altered conduction patterns, while also reducing computational demand. Since our electrophysiology simulations focus on this particular region, we developed automated Python scripts to crop each myocardium mesh at a manually identified plane, perpendicular to the long axis of the ventricles and identify appropriate boundary faces. This process produces meshed biventricular geometries, encompassing the left and right ventricles, epicardium (the outer surface of the heart), and base (the plane where the ventricles and atria connect). To standardize and refine the meshes, we utilize the meshing tools available in SimVascular, an open-source software package specifically designed for cardiovascular modeling and simulation \cite{simvascular}. By using SimVascular we are able to remesh each geometry in the cohort with a consistent resolution while also ensuring seamless integration into our simulation pipeline. 

\begin{figure}[H]
    \centering
    \begin{minipage}{0.49\textwidth}
        \centering
        \includegraphics[width=\textwidth]{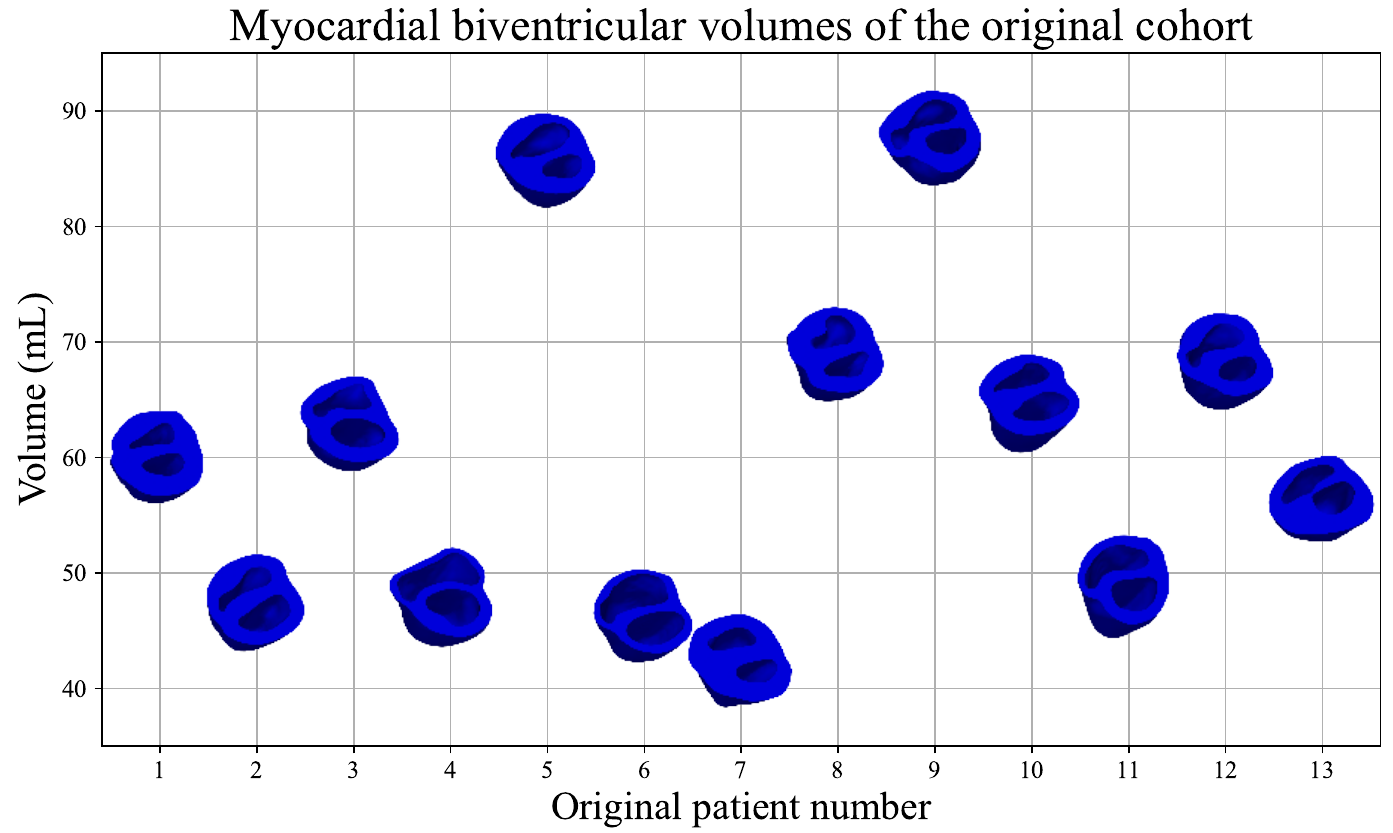}
    \end{minipage}%
    \hfill
    \begin{minipage}{0.49\textwidth}
        \centering
        \includegraphics[width=\textwidth]{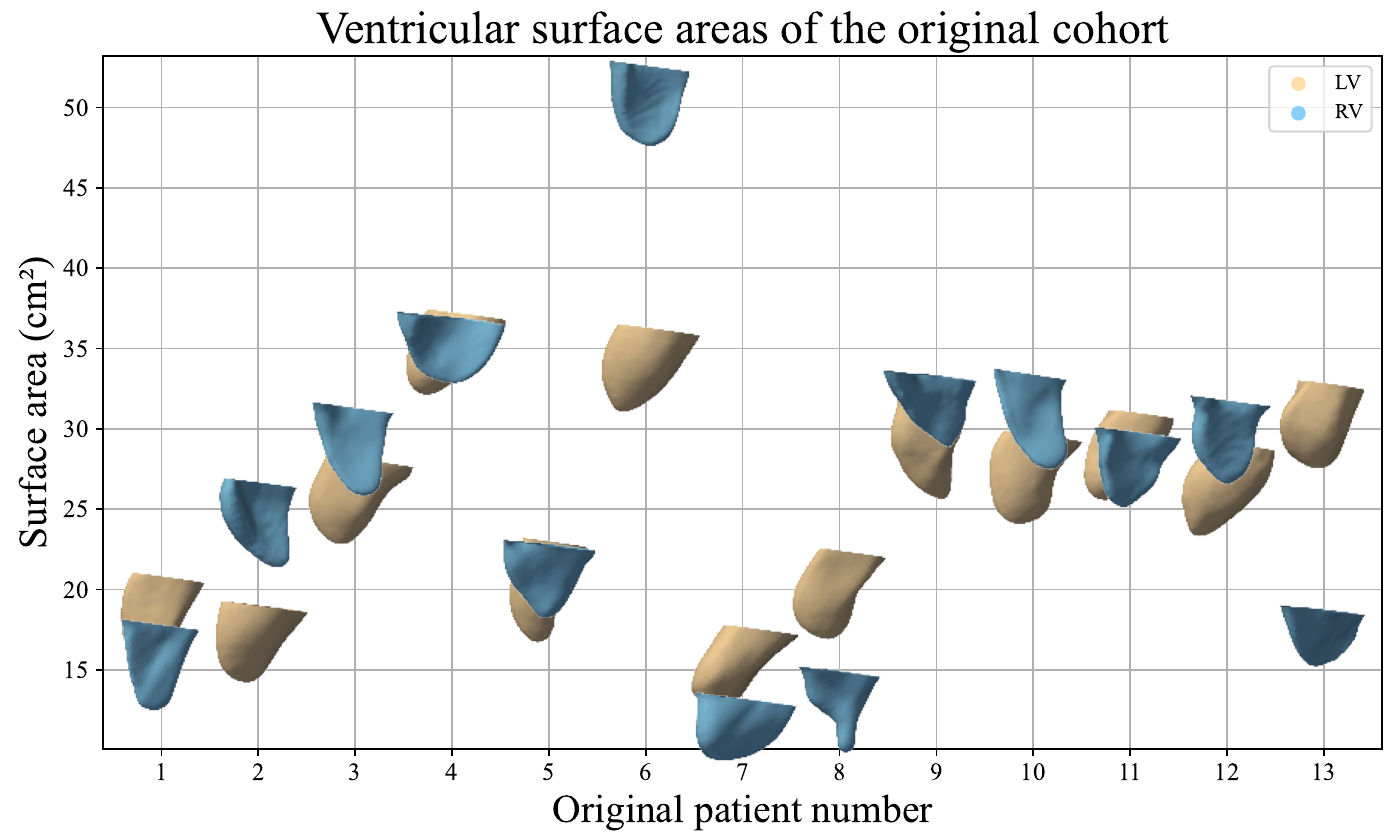}
    \end{minipage}

    \vspace{0.4cm} 
    
    \begin{minipage}{0.55\textwidth}
        \centering
        \includegraphics[width=\textwidth]{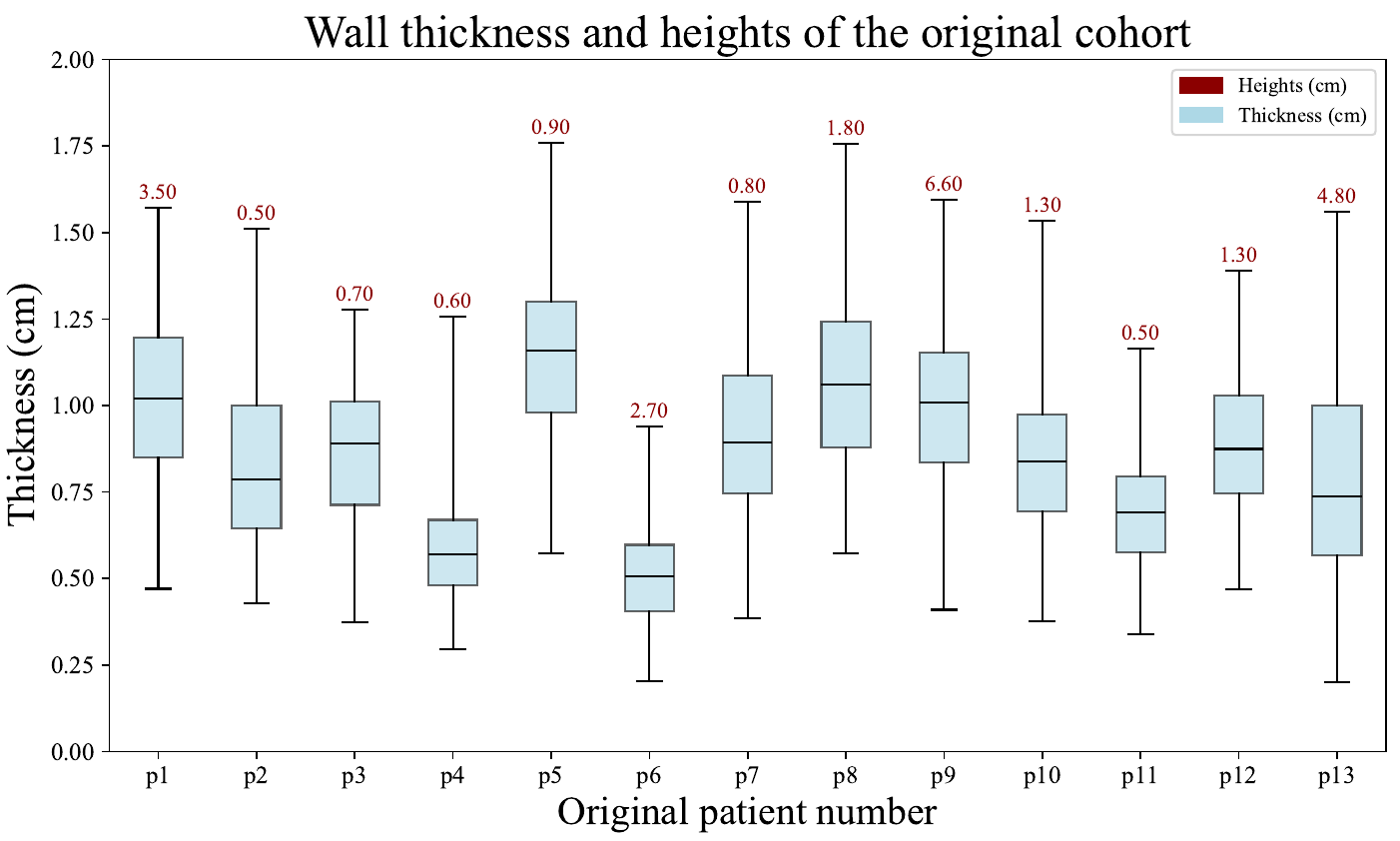}
    \end{minipage}
    
    \caption{Comparison of myocardial volumes, ventricular surface areas, and structural metrics (wall thickness and apicobasal heights) across our original patient cohort.}
     \label{combined_metrics}
     
\end{figure}

We study a small and challenging cohort of $\NumPatients =  13$ patients characterized by substantial anatomical variability. The myocardial tissue volumes across these biventricular geometries demonstrate substantial variation, with patient 5 exhibiting the highest volume at approximately 84.75 mL and patient 7 the lowest volume at 41.75 mL. We also observe significant variability in the endocardial surface areas of both ventricles.  which ranged from 15.80 cm$^2$ (patient 7) to 35.37 cm$^2$ (patient 4) for the left ventricle (LV), and from 12.04 cm$^2$ (patient 7) to 51.24 cm$^2$ (patient 6) for the right ventricle (RV). The largest difference in surface area between the LV and RV is 16.62 cm$^2$, while the smallest difference is 0.31 cm$^2$. In our cohort, the LV is larger in 5 out of 13 patients, whereas the RV is larger in the remaining 8 (Figure~\ref{combined_metrics}).

The myocardial wall thickness is defined as the shortest distance from any point on the endocardium (the surface of the LV and RV) to the nearest point on the epicardium, or the nearest point on the endocardium of the opposite ventricle when it is closer. This definition ensures that we accurately capture the thickness of the septum. We calculate the average wall thickness by averaging the shortest distances of all myocardial points in a geometry. Across our cohort of patients, the average wall thickness varies from 0.52 cm to 1.16 cm. Furthermore, we define the heights of the biventricular geometries as the distance from the apex (the tip of the epicardial surface) to its nearest point on the base. The biventricular heights vary across geometries, ranging from 0.50 cm to 6.6 cm. These measurements emphasize the substantial anatomical variability across patients in both wall thickness and apicobasal (from apex to base) height (Figure~\ref{combined_metrics}).

\subsection{Synthetic Cohort}
We extend the original cohort by combining anatomical mapping, geometrical encoding, and a novel sampling strategy to generate a synthetic cohort that is consistent with the characteristics of the original patient cohort described in the previous section. Figure \ref{fig:methods-SSM} presents an overview of our synthetic cohort generation pipeline.
\label{fig:methods-SSM}
\begin{figure}[h]
    \centering
    \includegraphics[width=1.0\linewidth]{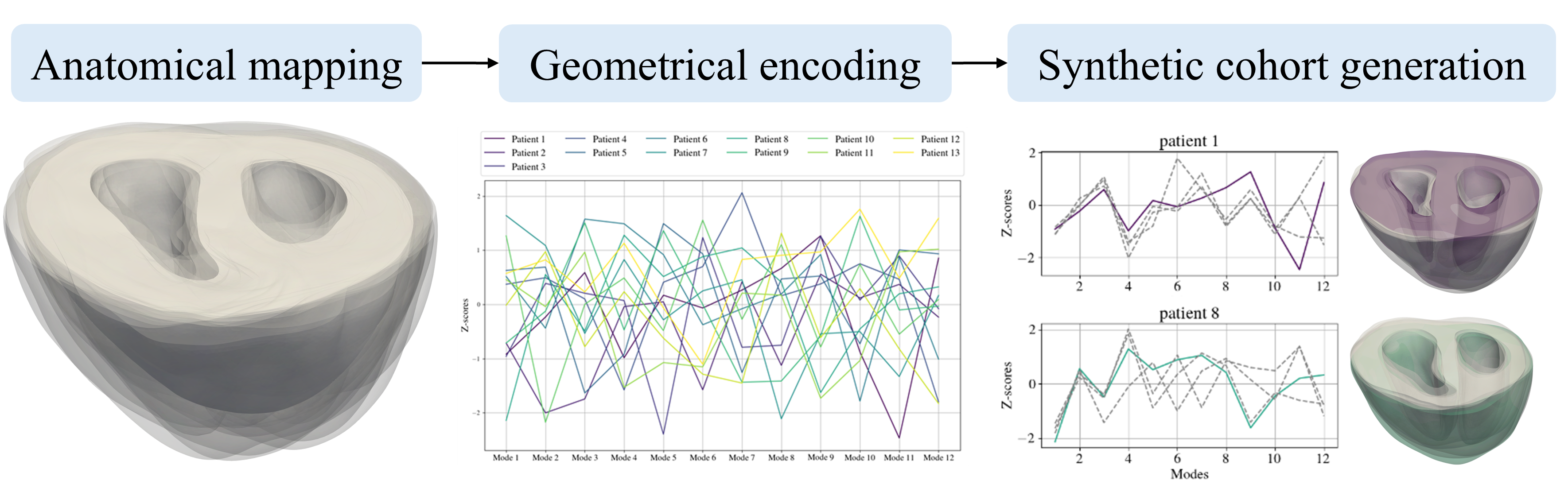}
    \caption{Overview of the synthetic cohort generation pipeline. We apply large deformation diffeomorphic metric mapping to map original anatomies onto a common template, followed by principal component analysis to encode the anatomical variability into z-scores. Using a novel sampling strategy, We generate synthetic anatomies based on their probabilistic proximity to the original samples in the multivariate space defined by these geometrical encodings. Specifically, we selected the 4 closest samples (dotted lines) to each original subject (colored lines).}
\end{figure}
\subsubsection{Large deformation diffeomorphic metric mapping}
\label{subsec:anatomical_mapping}
We perform mapping of the $\NumPatients = 13$ anatomies of the original cohort using the large deformation diffeomorphic metric mapping (LDDMM) approach detailed in Moscoloni et al. \cite{Moscoloni2025}. In this approach, we map a set of target patient-specific biventricular geometries to a template biventricular geometry by means of a dynamic flow of smooth, invertible transformations called diffeomorphisms. 
More specifically, these diffeomorphisms $\boldsymbol{\Phi}_i: \mathbb{R}^3 \mapsto \mathbb{R}^3$ entail a set of deformation vectors $\left( \boldsymbol{\mu}_n \right)_{n=1,\dots,N_q}$, known as momenta, which are applied to an ambient space of $N_q$ equally spaced control points $\left( \mathbf{q}_n \right)_{n=1,\dots,N_q}$.
We keep the initial set of control points fixed throughout the large deformation diffeomorphic metric mapping optimization procedure. Consequently, each target geometry defined as a surface mesh $\mathbf{\Gamma}_i$, is approximated by their unique set of momenta $\boldsymbol{\mu}_i$ deforming the template geometry $\mathbf{\Gamma}$:
\begin{equation}
\mathbf{\Gamma}_i \approx \boldsymbol{\Phi}_i \left( \mathbf{\Gamma} \right) = \boldsymbol{\Phi}_{\mathbf{q},\boldsymbol{\mu}_i} \left( \mathbf{\Gamma} \right)
\end{equation}
We construct diffeomorphisms $\boldsymbol{\Phi}_i$ by mapping each point from an initial template geometry $\mathbf{\Gamma}$ to a corresponding point on a target geometry $\mathbf{\Gamma}_i$. This mapping is defined through a spatiotemporally varying velocity field $\mathbf{v}(\mathbf{x},t) = \mathbf{v}_t(\mathbf{x})$ over a \textit{pseudo-time} $t \in [0,1]$ \cite{glaunès_2008_lddmm_root,Moscoloni2025}.
We express the velocity field explicitly at each space location $\mathbf{x}$ and  time $t$ as: 
\begin{equation}
\mathbf{v}_t (\mathbf{x}) = \sum_{k=1}^{N_q} K_V \left( \mathbf{x}, \mathbf{q}_k(t) \right){\boldsymbol\mu}_k(t),
\label{eqn:vel_field_point}
\end{equation}
where $K_V$ is a Gaussian kernel $K_V\left( \mathbf{x}, \mathbf{y} \right) = \exp \left(- \|\mathbf{x} - \mathbf{y}\|^2 / {\lambda_V}^2 \right)$ of kernel width $\lambda_V$. This kernel weighs the contribution of momenta $\boldsymbol{\mu}_k(t)$ applied at control points $\mathbf{q}_k(t)$ to the instantaneous movement of the template vertex at position $\mathbf{x}$. 

\medskip \noindent
The control points themselves evolve dynamically under the same velocity field (Eq. \ref{eqn:vel_field_point}). Their trajectories follow Hamiltonian equations of motion \cite{deformetrica_2018}, governed explicitly by their momenta $\boldsymbol{\mu}(t)$:
\begin{equation}
\left\{\begin{array}{l}
\dot{\mathbf{q}}(t)=K_V(\mathbf{q}(t), \mathbf{q}(t)) \cdot \boldsymbol{\mu}(t) \\
\dot{\boldsymbol{\mu}}(t)=-\frac{1}{2} \nabla_\mathbf{q}\left\{K_V(\mathbf{q}(t), \mathbf{q}(t)) \cdot \boldsymbol{\mu}(t)^{\top} \boldsymbol{\mu}(t)\right\}
\end{array}\right.
\label{eqn:time_varying_momenta}
\end{equation}

\medskip \noindent
Given an initial set of control points $\mathbf{q}(0)$, we select time-varying momenta $\boldsymbol{\mu}(t)$ that define deformation paths $\boldsymbol{\Phi}_t$ that bring the template geometry as close to the target geometry of interest as possible at $t=1$.
Since multiple possible deformations paths exist, we prioritize paths that minimize the kinetic energy, resulting in geodesic trajectories. These geodesics are fully parametrized by the initial momenta $\boldsymbol{\mu}_0=\boldsymbol{\mu}(0)$ \cite{Durrleman2014,Moscoloni2025}.

\medskip \noindent
To estimate the optimal template geometry $\mathbf{\Gamma}$ and deformation fields $\boldsymbol{\Phi}_{\mathbf{q},\boldsymbol{\mu}_i}$, we minimizing the following cost function with respect to the initial template geometry and the set of initial momenta:
\begin{equation}
\mathcal{C} \left( \mathbf{\Gamma} , \boldsymbol{\mu} \right) = 
\sum_{i=1}^{N_\mathcal{O}} \frac{1}{\sigma^2} d_W\left(\boldsymbol{\Phi}_{\mathbf{q},\boldsymbol{\mu}_i} \left( \mathbf{\Gamma} \right),\mathbf{\Gamma}_i\right)^2 
+ \sum_{i=1}^{N_\mathcal{)}} \boldsymbol{\mu}_{i}^\top {K}_W(\mathbf{q}, \mathbf{q})\boldsymbol{\mu}_{i},
\label{eqn:cost_function_mapping}
\end{equation}
where the first term minimizes the sum of the Varifold distances $d_W$ between the target geometries $\mathbf{\Gamma_i}$ and their approximations through $\boldsymbol{\Phi}_{\mathbf{q},\boldsymbol{\mu}_i}$, while the second term penalizes deformations with high kinetic energy. Here, $\sigma$ balances the relative contributions of these two terms.  

\medskip \noindent
We optimize Eq. \ref{eqn:cost_function_mapping} iteratively through two alternating steps: 
(i) We first optimize the momenta $\boldsymbol{\mu}$ for each target geometry,  keeping the template fixed $\boldsymbol{\Gamma}$. This step identifies the best deformation momenta to match the current)template to each individual target geometry. 
(ii) Next, we optimize the template $\boldsymbol{\Gamma}$ itself, while keeping the previously obtained momenta $\boldsymbol{\mu}$ fixed. This step updates the template to better represent the average of the current approximations of the target geometries derived from the current set of subject-specific momenta $\boldsymbol{\mu}$. 
We employ a gradient descent scheme for both optimization steps, and repeat these steps until convergence of the initial momenta and template is achieved \cite{deformetrica_2018, glaunès_2008_lddmm_root,Moscoloni2025}. 

\medskip \noindent
After optimization, the final estimated template geometry $\mathbf{\Gamma}$ along with subject-specific momenta $\boldsymbol{\mu_i}$ for each subject provides a complete mapping of each subject's anatomy within the cohort. 
Applying these subject-specific momenta to the vertices of the optimized template mesh explicitly reconstructed individual anatomical surface meshes. 
In practice, this mapping process entails the selection of an initial template, to which the surface meshes need to be aligned, and two user-defined hyperparameters that influence the accuracy of the mapping: the Gaussian kernel width of $\lambda_V$ from Eq. (\ref{eqn:time_varying_momenta}), and the grid spacing parameter $\lambda_W$ defining the density of the control point grid. Large $\lambda_V$ values promote global, smooth deformations, while large $\lambda_W$ values allow capturing fine anatomical details within target meshes \cite{bruse_2016_ssm_aorta}.

\subsubsection{Statistical shape modeling pipeline}
\textbf{Mesh alignment and anatomical mapping}.
To apply the LDDMM strategy outlined in the previous section to our cohort, we first align the target meshes and extract an initial template geometry. Starting with a cohort of $\NumPatients$ surface meshes, we align the meshes in two steps. 
First, we align each mesh to an initial reference mesh, chosen as the mesh with the lowest number of vertices, using the iterative closest point (ICP) algorithm between each mesh and the initial target \cite{besl_mckay_1992_ICP}. Briefly, ICP  finds the optimal rigid body transformation minimizing the distances between corresponding points based on spatial proximity. Following this preliminary alignment, we select the medoid mesh - the instance which closest to the cohort's mean - as our updated reference mesh and perform a second ICP alignment to this updated reference mesh, which subsequentlu serves as the initial template for the anatomical mapping.
With this initial template selected, we perform LDDMM anatomical mapping using Deformetrica \cite{deformetrica_2018}. We optimize two hyperparameters, related to the Gaussian kernel width $\lambda_V$ and the control point grid resolution $\lambda_W$ (see section \ref{subsec:anatomical_mapping}. Following hyperparameter tuning, we set $\lambda_V$ = $\lambda_W $= 6 mm. This hyperparameter choice yields a total of 1100 control points, which adequately balances global deformation and local anatomic detail. 
The initial template, control points, and model hyperparameters are used to estimate the final template mesh and patient-specific momenta for the cohort. The combination of control points, optimized template, and patient-specific momenta allows us to reconstruct each patient-specific geometry with consistent point-to-point correspondence. 

\medskip \noindent
\textbf{Evaluating reconstruction accuracy}.
We quantify the accurarcy of our anatomical mapping through reconstruction errors, measured as the average vertex-to-vertex distance between original mesh and reconstructed meshes. Specifically, let $\left( \mathbf{p} \right)_{j = 1,\dots,N_v}$ be the set of vertices in the original mesh for patient $i$, , and $\left( \mathbf{r} \right)_{j = 1,\dots,N_v}$ be the corresponding vertices in the reconstructed mesh, where $N_v$ is the total number of points. We define the reconstruction error $\ell_i$ for patient $i$ as:
\begin{equation}
\ell_i = \frac{1}{N} \sum_{j=1}^{N_v} \|\mathbf{p}(j)_i - \mathbf{r}(j)_i\|
\end{equation}
where $\|\mathbf{p}(j)_i - \mathbf{r}(j)_i\|$ is the Euclidean distance between the $j$-th vertex point on the original mesh and the $j$-th vertex point on the reconstructed mesh for patient $i$. 
The overall reconstruction accuracy for the cohort of $\NumPatients$ patients is quantified by the average total reconstruction error $\ell_{\text{tot}}$ across all individuals:
\begin{equation}
\ell_{\text{tot}} = \frac{1}{\NumPatients} \sum_{i=1}^{\NumPatients} \ell_i
\end{equation}

\medskip \noindent
\textbf{Principal component analysis and shape encoding}.
To capture the anatomical variability of the cohort, we perform principal component analysis (PCA) on the spatial coordinates of the corresponding vertices across the cohort \cite{Moscoloni2025, kollar_2022_rv_ssm}. PCA decomposes the geometry variations into a set of principal components, allowing each patient's geometry to be represented compactly as:
\begin{equation}
\mathbf{\Gamma}_i = \bar{\mathbf{\Gamma}} + \sum_{m=1}^{\NumModes} \alpha_{i,m} \boldsymbol{\phi}_m \lambda_m, \quad i \in \{1, 2, \ldots, \NumPatients\},
\label{eqn:linear_comb_shape_scores}
\end{equation}
where $\phi_m$ is the $m^{th}$ principal component, and $\alpha_{i,m}$ is the patient specific weight, or shape score, of that component in the $\textit{i}^{\text{th}}$ patient.
This PCA-based representation allows efficient encoding of anatomical variability using only the shape scores $\alpha_{i,m}$. We found that using 12 principal component-derived z-scores sufficiently captures the anatomical variability within the patient cohort. For the sampling and training of the surrogate model (section \ref{sec:methods:BLNM}), we normalized the shape coefficients as z-scores by their mean and standard deviation.
Furthermore, we can generate new synthetic patient geometries by sampling novel shapes scores $\beta_{i,n}$ and plugging them into Eq. \eqref{eqn:linear_comb_shape_scores} \cite{bruse_2016_ssm_aorta, romero_2021_synthetic_gen}. 

\subsubsection{Synthetic cohort generation}
To augment our original cohort of $\NumPatients$ anatomies with additional synthetic yet realistic data, we introduce a novel three-step pipeline based on geometrical encoding and probabilistic sampling.
This synthetic cohort generation procedure includes: 
(i) sampling new candidate shape scores and filtering them based on the probability density function of the original cohort, 
(ii) identifying candidate shape scores that are the closest to the original patients in the multivariate space,
(iii) reconstructing and meshing synthetic anatomies from selected shape scores

\medskip \noindent
\textbf{Probabilistic sampling and filtering}.
We employ Latin Hypercube Sampling (LHS) \cite{Peirlinck2019}  to generate synthetic candidate shape scores within a 12-dimensional multivariate space , with values ranging from [-2.5 , 2.5] standard deviations in each mode of anatomical variation.  We sample a large set of samples to ensure broad coverage of anatomical variations. 
To ensure compatibility of synthetic anatomies with our original cohort, we filter these candidate shape scores based on their probability relative to the original cohort of $\NumPatients$ anatomies. 
Specifically, we estimate the underlying probability density function using a non-parametric Gaussian Kernel Density Estimation (KDE) \cite{chen_2017_kde} applied to the original z-scores: 
\begin{equation}
\hat{f}(\boldsymbol{z}) = \frac{1}{n h_1 h_2 \ldots h_d} \sum_{i=1}^n K\left(\frac{\boldsymbol{z} -\boldsymbol{Z}_i}{h}\right),
\end{equation}
where $\hat{f}(\boldsymbol{z})$ is the estimated density function at point $\boldsymbol{z}$,  $n$ is the number of data points in the original cohort, \( \boldsymbol{Z}_i \) are the sample points from the original data, \( h_1, h_2, \ldots, h_d \) are the bandwidths for each dimension, determining the width of the kernel, and \( K(u) \) is the Gaussian kernel function.

\medskip \noindent
This method allows us to model the distribution of the original z-scores without assuming a specific parametric form. We apply this KDE to the z-scores derived from the original cohort of $\NumPatients$ anatomies, to derive the probability distribution across the multivariate space defined by the modes of variation. Using this estimated density, we define a threshold probability corresponding to the least probable sample in the original cohort. We filter out any LHS-generated candidate scores falling below this threshold, thus ensuring that our synthetic cohort accurately reflects realistic anatomical variations observed in our original population.

\medskip \noindent
\textbf{Selecting closest synthetic samples}.
Following probabilistic filtering, we identify synthetic generated samples that closely resemble actual patient geometries. We  first transform both the original and synthetic normalized z-scores back to their original scales using the respective means and standard deviations of each mode. This approach ensures that the distances between shape scores reflect actual anatomical variations appropriately.

\medskip \noindent
We then compute the Euclidean distances in the multivariate space between each pair of original shape scores $\alpha_{i, n}^{\text {orig}}$ and the filtered synthetic shape scores $\beta_{j, n}^{\text {gen}}$ : 
\begin{equation}
\ell_{i j}=\sqrt{\sum_{n=1}^N\left(\alpha_{i, n}^{\text {orig}}-\beta_{j, n}^{\text {gen}}\right)^2}
\end{equation}
where $N$ is the total number of modes considered.
For each original patient, we rank these synthetic candidates by their proximity, selecting the four closest matches. 
This step enables us to focus further on the samples that fall not only within the statistically probable range but are also geometrically close to the original data points, thus effectively enhancing the diversity yet maintaining anatomical realism in the synthetic cohort. 

\medskip \noindent
\textbf{Reconstruction and mesh generation}.
We reconstruct synthetic surface meshes from the selected shape scores by applying Equation (\ref{eqn:linear_comb_shape_scores}), resulting in a cohort of $\syntheticPatients = 52$ synthetic surface meshes. 
These surface meshes are further processed using Meshmixer \cite{schmidt_singh_2010_meshmixer} to fix small topological defects that may arise during reconstruction. 
We subsequently generate volumetric meshes for each synthetic patient using Gmsh \cite{geuzaine_remacle_2009_gmsh}. 
Finally, to ensure full compatibility with our original cohort and integration with our simulation pipeline, we remesh all synthetic volumetric geometries using SimVascular. 

\subsection{Cardiac electrophysiology}
\label{sec:methods:EP}

We introduce the mathematical model and the numerical scheme that we adopt to generate activation sequences of the biventricular geometries. We simulate cardiac electrophysiology for both the synthetic and original patient cohorts by generating a total of $\totalPatients = 65$ activation maps.

\subsubsection{Mathematical model}
\label{sec:methods:EP:model}

We model cardiac electrophysiology in multiple 3D domains $\Omega^i = \Omega_\mathrm{BiV}^i \cup \Omega_\mathrm{purk}^i \subset \mathbb{R}^3$ represented by biventricular-Purkinje models of original and synthetic patients affected by ToF using the monodomain equation \cite{collifranzone2014book,Quarteroni2019,Peirlinck2021a} coupled with the ten Tusscher-Panfilov ionic model \cite{TTP06}, that is:

\begin{eqnarray} \label{eqn:monodomain}
\left\{\begin{array}{ll}
\displaystyle
\frac{\partial \Pot}{\partial t}+\Iion(\Pot,\Ionic)
-\nabla\cdot(\DiffTens \nabla \Pot)=\Iapp({\bf x},t) & \mbox{ in }\Omega^i \times (0,T],\\[2mm]
(\DiffTens\nabla \Pot)\cdot {\bf n}=0  & \mbox{ on }\partial\Omega^i \times (0,T],\\[2mm]
\displaystyle
\frac{d\Ionic}{dt}=\RhsIonic(\Pot,\Ionic)
& \mbox{ in }\Omega^i \times (0,T],\\[2mm]
\displaystyle
\Pot({\bf x},0)=\Pot_0({\bf x}),\
\Ionic({\bf x},0)=\Ionic_0({\bf x}) &  \mbox{ in } \Omega^i,
\end{array}\right.
\end{eqnarray}

where $i = 1, ..., \totalPatients$, where  $\totalPatients = \NumPatients + \syntheticPatients$ represents the total count of biventricular geometries in the combined original and synthetic cohorts. The primal variable $\Pot$ defines the transmembrane potential, that is the electric signal propagating throughout the cardiac tissue, whereas vector $\Ionic=(y_1, \ldots, y_{M+P})$ introduces both the probability density functions of $M=12$ gating variables, which represent the fraction of open channels across the membrane of a single cardiomyocyte, and the concentrations of $P=6$ relevant ionic species, such as intracellular calcium $Ca^{2+}$, sodium $Na^{+}$ and potassium $K^{+}$.

The right hand side $\RhsIonic(\Pot,\Ionic)$ defines a system of ordinary differential equations for the time evolution of gating and concentration variables.
The ionic current $\Iion(\Pot,\Ionic)$ enables cell-to-organ scale coupling between the ionic and monodomain models.
The analytical expressions of both $\RhsIonic(\Pot, \Ionic)$ and $\Iion(\Pot, \Ionic)$ derive from the mathematical formulation of the ten Tusscher-Panfilov ionic model \cite{TTP06}. The diffusion tensor is expressed as $\DiffTens = \Di \IdentityVec + \Da \fZero \otimes \fZero$ in $\Omega_\mathrm{BiV}$ and $\DiffTens = \Dp \IdentityVec$ in $\Omega_\mathrm{purk}$, where $\fZero$ defines the biventricular fiber field \cite{SahliCostabal2018,Peirlinck2021a,Piersanti2021}.
$\Da, \Di, \Dp \in \mathbb{R}^+$ represent the anisotropic, isotropic and Purkinje conductivities, respectively \cite{Peirlinck2022}. 
We impose homogeneous Neumann boundary conditions on $\partial\Omega^i$, where $\boldsymbol{n}$ is the outward unit normal vector to the boundary, to prescribe an electrically isolated domain.

The action potential is triggered by a current $\Iapp({\bf x},t)$ applied at time $t = 0$ at the first nodes of both the left and right bundle branches. These nodes initiate the depolarization of the right and left ventricle Purkinje networks, which we generate using a fractal tree algorithm. This algorithm effectively covers the complex and irregular surfaces inside the ventricles through a two-step projection process \cite{Costabal2016}. We construct a fractal network by defining branch parameters, including initial length, generation count, median length, length standard deviation, angle, and a repulsion parameter which regulates curvature. Currently, no in vivo imaging technique can accurately capture Purkinje networks \cite{Alvarez2025}, which exhibit both significant inter- and intrapatient variability in size and shape. We utilize a coverage-based approach to achieve a consistent and reproducible method to automatically create Purkinje networks across geometries.

\begin{figure}[h]
  \centering
  \includegraphics[width=.75\textwidth]{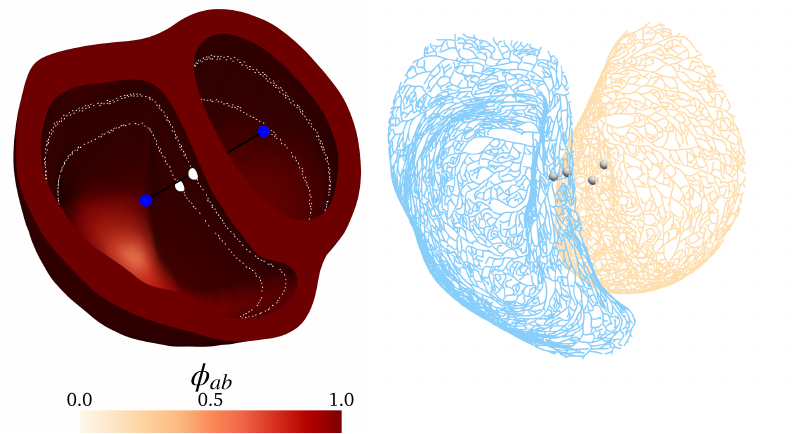}
   \caption{Selection of the initial points for the Purkinje networks. On the left, the white contour points represent universal ventricular coordinates, $\phi_{ab} = 0.97$ and $\phi_{ab} = 0.99$, derived from the apex-to-base simulation of the Laplace heat equation \cite{Bayer2012}. 
   The centroids of the $\phi_{ab}$ points for the LV and RV are marked by blue spheres. The first and second points in each ventricle are represented by white spheres, where the second point is chosen to determine the direction of the first segment. On the right, the corresponding RV and LV Purkinje networks are visualized in light blue and orange, respectively}
      \label{Purkinje}
\end{figure}

As a first step in this approach, we methodically select two points in each ventricle to determine the direction of the first branch of each Purkinje network, as shown in Figure~\ref{Purkinje}. This selection is guided by heat simulations previously conducted to activate different regions of the biventricular geometries. From the apex-to-base heat simulation, we extract all points where the universal ventricular coordinate $\phi_{ab}$ reaches values of 0.99 and 0.97 for each ventricle. Next, we calculate the centroids of all the $\phi_{ab} = 0.99$ points for the LV and RV separately and draw a line connecting these two centroids. The starting points for the left and right Purkinje networks are then identified as the $\phi_{ab} = 0.99$ points closest to this line in the LV and RV respectively. For each network, the second point is designated as the $\phi_{ab} = 0.97$ point closest to its respective initial point. This second point does not necessarily correspond to the second node in the network but rather serves to determine the direction of the first segment. This systemic approach, ensures that the selected points are geometrically aligned and standardizes the process of determining the direction of the initial branch for each electrophysiology simulation.

For all geometries, we set median branch length $l = 1.8$ mm, standard deviation $\sigma = \sqrt{0.2}*l$, branch angle $\alpha = 0.2$ radians, and branch repulsion $w = 0.15$. We iterate the initial length and branch generation count, starting at 6.5 mm and 22 generations, respectively, until our algorithm produces a Purkinje network where at least $82\%$ of all endocardial points in the ventricle are no more than 0.5 mm away from a node in the created network.

In the absence of ECG data for our geometries, we adopt parameter values from a previous study offering representative ECG profiles for CHD cases \cite{Salvador2024BLNM}.
We apply a stimulus with amplitude -52.0 $\mu$A/cm$^2$ to the initial nodes of each of the left and right Purkinje networks for 5.0 ms and a cycle length of 800.0 ms. We conduct all electrophysiology simulations using a 3D monodomain model with biventricular and Purkinje fiber domains. Simulations are performed for each geometry in our original and synthetic patient cohorts using svMultiPhysics, a high-performance C++ solver in the SimVascular project designed for multi-physics and multi-scale finite element simulations of the cardiovascular system \cite{Zhu2022}. The workflow from running heat simulations to generating the Purkinje network is performed on a single CPU. For electrophysiology simulations, we utilize a single node from the Stanford Research Computing Center cluster per simulation, with each node featuring 24 cores. Table \ref{tab:ep_parameters} summarizes the key parameters used in the simulations, including conductance values for ionic currents and domain-specific tissue conductivities.

\begin{table}[h]
\centering
\begin{tabular}{| c | c | c |}
\hline
\textbf{Parameter} & \textbf{Value} & \textbf{Units} \\ \hline
Maximal current conductance of $Na^+$ & 13.19 & nS pF$^{-1}$ \\ \hline 
Maximal current conductance of $Ca^{2+}$ & 7.91e-05 & cm ms$^{-1}$ $\mu$F$^{-1}$ \\ \hline
Maximal current conductance of the rapid delayed rectifier $I_{Kr}$ & 0.11 & nS pF$^{-1}$ \\ \hline
Myocardial anisotropic conductivity & 0.0177 & mm$^2$ ms$^{-1}$ \\ \hline
Myocardial isotropic conductivity & 0.0059 & mm$^2$ ms$^{-1}$ \\ \hline
Purkinje fiber anisotropic conductivity & 1.60 & mm$^2$ ms$^{-1}$ \\ \hline
Purkinje fiber isotropic conductivity & 1.60 & mm$^2$ ms$^{-1}$ \\ \hline
\end{tabular}
\caption{Cardiac electrophysiology simulation parameters.}
\label{tab:ep_parameters}
\end{table}

\subsubsection{Numerical discretization}
\label{sec:methods:EP:discretization}

We perform space discretization of the monodomain equation coupled with the ten Tusscher-Panfilov ionic model using the Finite Element Method (FEM) with $\mathbb{P}_1$ Finite Elements.
The tetrahedral mesh of the biventricular-Purkinje system has an average mesh size $h = 1$ mm, for each patient.
We apply non-Gaussian quadrature rules to recover convergent conduction velocities of the biventricular-Purkinje system \cite{Tikenogullari2023}.

For time discretization, we employ an Implicit-Explicit numerical scheme where we first update the variables of the ionic model and then the transmembrane potential \cite{Regazzoni2022,Piersanti2022,Fedele2023}.
In particular, the diffusion term in the monodomain equation is treated implicitly and the ionic term is treated explicitly.
Furthermore, we use the ionic current discretization by means of the Ionic Current Interpolation scheme \cite{Krishnamoorthi2013}.
We employ the Forward Euler scheme with a time step size $\Delta t=0.1$ ms and we simulate ventricular depolarization only by considering $T= 100$ ms, for each patient.

We apply a Laplace-Dirichlet Rule-Based algorithm \cite{Bayer2012} to generate the fiber, sheet and sheet-normal distributions on the myocardium using the following parameters: $\alpha_\mathrm{epi}$ = $-60^\circ$, $\alpha_\mathrm{endo}$ = $60^\circ$, $\beta_\mathrm{epi}$ = $20^\circ$ and $\beta_\mathrm{endo}$ = $-20^\circ$.

\subsubsection{Activation Maps}
\label{sec:methods:EP:activationMaps}
We generate activation maps from simulation outputs consisting of 1000 VTK files, each representing a 0.1 ms time step over a total duration of 100 ms. This fine temporal resolution is essential to accurately capture the rapid propagation of the cardiac action potential, which occurs over just a few milliseconds. By sampling at 0.1 ms intervals, we ensure sufficient temporal resolution to identify precise activation patterns. We process each file sequentially to extract the action potential at each node in the mesh. As soon as the action potential of a node exceeds the depolarization threshold of -10 mV, we record the corresponding simulation time as the activation time for that node. This depolarization threshold indicates the physiological transition from resting to active. We then construct an activation map by applying these activation times as scalar values to the 3D cardiac model. An individual activation map is generated for each original and synthetic patient, i.e. 65 activation maps in total.

\subsection{Branched Latent Neural Maps}
\label{sec:methods:BLNM}

We construct a surrogate model of cardiac function that generalizes across different anatomies by building a feedforward partially-connected neural network with Tanh activation functions. The model is trained to learn activation maps post-processed from the numerical solution of our physics-based electrophysiology model while structurally separating the role of the space coordinates $\mathbf{x}$ and the anatomical parameters $\paramBiV^i$. This recently proposed scientific machine learning tool \cite{Salvador2024BLNM} allows for different levels of disentanglement between inputs and outputs.
The surrogate model reads:
\begin{equation} \label{eqn:BLNM}
    \ANNState(\mathbf{x}) = \ANNRhs \left(\mathbf{x}, \paramBiV^i; \ANNparam \right) \text{ for } \mathbf{x} \in \Omega^i, \; i \in \{1, ..., \totalPatients\}.
\end{equation}

The set $\ANNparam \in \mathbb{R}^{\NumANNWeights}$ encodes weights and biases of a feedforward partially-connected NN, which represents a map $\ANNRhs \colon \mathbb{R}^{3 + \NumModes} \to \mathbb{R}^{\NumANNState}$ from space coordinates $\mathbf{x} \in \mathbb{R}^3$ and model parameters $\paramBiV^i \in \paramSpace \subset \mathbb{R}^{\NumModes}$ to an output vector $\ANNState(t) = [\ANNStateAT(t), \ANNLatent(t)]^T \in \mathbb{R}^{\NumANNState}$. In this application, the vector $\paramBiV$ represents the anatomical z-scores of a biventricular geometry. When the model utilizes all 12 extracted anatomical z-scores, $\NumModes= 12$ and the vector is defined as $\paramBiV = [\theta_1, \theta_2, . . . , \theta_{12}]$.

$\ANNState(t)$ contains activation times $\ANNStateAT(\mathbf{x}) \in \mathbb{R}$ on the different geometries $\Omega^i$. Furthermore, the vector leverages some $\ANNLatent(t)$ latent variables that enhance the learned space dynamics by locally acting in regions with sharp features and steep gradients. The presence of partial connections contributes to a smoother  landscape of local minima compared to fully-connected neural networks.
For this reason, we optimize neural network parameters $\ANNparam$ by using the Adam optimizer in place of a second-order optimizer \cite{Salvador2024BLNM}. This allows for faster training while accounting for similar performance in accuracy with respect to more complex optimizers \cite{Salvador2024BLNM}.

During nonlinear optimization, we track the mean squared error (MSE) produced by the surrogate model with respect to physics-based numerical simulations to find an optimal set of weights and biases $\ANNparam$, that is:
\begin{equation} \label{eqn:loss_BLNM}
\mathcal{L}(\ANNStateATAdim(\widetilde{\mathbf{x}}), \ANNStateTildeObs(\widetilde{\mathbf{x}}); \ANNparamTrained) = \underset{\ANNparamTrained}{\arg\min} \left[ || \ANNStateATAdim(\widetilde{\mathbf{x}}) - \ANNStateTildeObs(\widetilde{\mathbf{x}}) ||_{\text{L}^2(\Omega^i)}^2 \right] \text{ for } i \in \{ 1, ..., \totalPatients \}.
\end{equation}
where $\ANNStateATAdim(t) \in [-1, 1]$ represents BLNM outputs and $\ANNStateTildeObs(t) \in [-1, 1]$ defines the post-processed activation maps from physics-based numerical simulations, both in non-dimensional form.
Space $\widetilde{\mathbf{x}} \in [-1, 1]^3$ and z-scores modes $\paramBiVAdim^i \in [-1, 1]^{\NumModes}$ are also normalized during the training and inference.
For further details on the BLNM method, we refer the reader to \cite{Salvador2024BLNM}. Furthermore, the Python implementation of our BLNM model is publicly available under MIT License at \url{https://github.com/StanfordCBCL/BLNM}.

\subsubsection{Preparation of Training and Testing Sets}
We generated activation maps for 52 synthetic and 13 actual ToF patient geometries. Across our 65 geometries, sizes vary significantly, with the smallest biventricular mesh containing 166,026 points and the largest 372,117 points. The average number of points per geometry is approximately 265,394, with a standard deviation of 42,033 points. Prior to training, we normalize both spatial coordinates and z-scores for each geometry to ensure consistency across different scales. Our training set consists of synthetic geometries, while our testing set contains the original ToF patient cohort.

\subsubsection{BLNM Hyperparameter Tuning}
We use Ray Tune, a Python library designed to optimize the search for the best model hyperparameters. We explore 100 neural network configurations, where each configuration is generated by selecting hyperparameters within predefined ranges. Each hyperparameter, except for the learning rate, is sampled from a discrete set of values. These ranges are listed in the Potential Values column of Table 2. The learning rate is sampled from a logarithmic uniform distribution between $10^{-4}$ and $10^{-3}$. We employ OptunaSearch to efficiently identify the hyperparameters that yield the lowest MSE. OptunaSearch is a hyperparameter optimization algorithm within Ray Tune that leverages Optuna, a flexible and efficient optimization library \cite{optuna}. This sampling process is repeated throughout the 1,000-epoch training period, with model performance evaluated every ten epochs. 

\begin{figure}[h]
  \centering
  \includegraphics[width=1\textwidth]{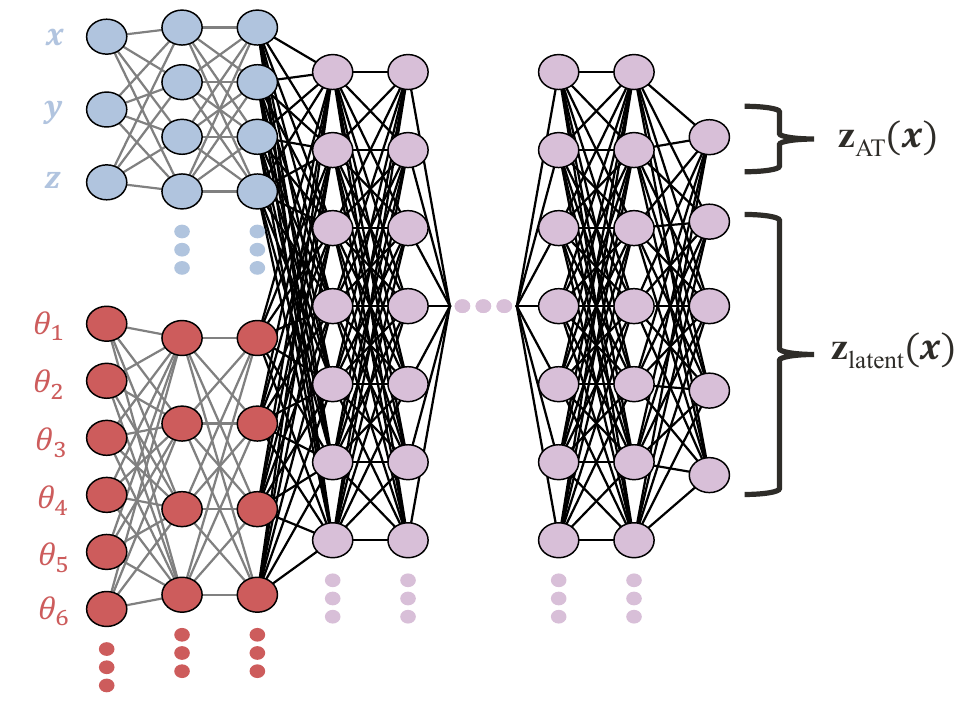}
   \caption{Sketch of the optimal BLNM architecture of 16 layers with 42 neurons per layer and 5 states as determined using Ray Tune with the OptunaSearch algorithm. The network starts with separate branches: the first branch processes spatial coordinates (x, y, z), and the second processes the geometry-specific modes $\theta_1$ through $\theta_{12}$. These branches operate independently up to a specified disentanglement level of 2 before merging. }
   \label{blnm}
\end{figure}

\begin{table}[h]
\centering
\begin{tabular}{|c | c | c |}
\hline
\textbf{BLNM Parameters} & \textbf{Potential Values} & \textbf{Optimal Values }\\
\hline
Layers & [5, 500] & 16  \\ \hline
Neurons & [5, 20] & 42 \\ \hline
Learning Rate & logU($10^{-4}, 10^{-3}$) & 0.0005 \\ \hline
\# States & [1, 5] & 5 \\ \hline
Disentanglement & [1, 4] & 2 \\ \hline
\# of Training Points & [1500, 10000] & 1485 \\ \hline
\end{tabular}
\caption{Ranges and Ray Tune-optimized parameter values for BLNM.}
\label{tab:blnms}
\end{table}
The optimal BLNM configuration consists of 16 layers with 42 neurons per layer, 5 states (3 spatial parameters, 12 anatomical variability modes, and 4 latent variables), and a disentanglement level of 2 (Table 2). Fig.~\ref{blnm} provides a visualization of this structure.  During each epoch, we randomly select 1,485 points from our training geometries as spatial coordinates. Model performance is evaluated every ten epochs, with the entire training process spanning a total of 1,000 epochs.

	\section{Results}
\label{sec:results}
We present the mapping results, including the reconstruction accuracy achieved after our anatomical mapping. We also present the activation maps generated from our cardiac electrophysiology simulations for both the original and synthetic cohort. Finally, we showcase the results of our BLNM model trained on the synthetic cohort. Our findings demonstrate that BLNMs can accurately predict cardiac activation times across ToF geometries with substantial anatomical variability, achieving an average mean squared error of 0.0034 on the testing set. 

\subsection{Anatomical mapping results and synthetic cohort}
Our anatomical mapping process yields an average reconstruction error of 0.086 mm across the cohort, with individual patient reconstruction errors ranging from  0.055 to 0.11 mm. This low error indicates a high accuracy in capturing subject-specific anatomies using our methodology. We then apply PCA to the mapped geometries, choosing a finite set of $\NumModes$ = 12 principal modes of variation. This choice effectively captures essential anatomical features while reducing dimensional complexity, which facilitates efficient sampling and subsequent analysis.
\begin{figure}[h]
    \centering
    \includegraphics[width=1.0\linewidth]{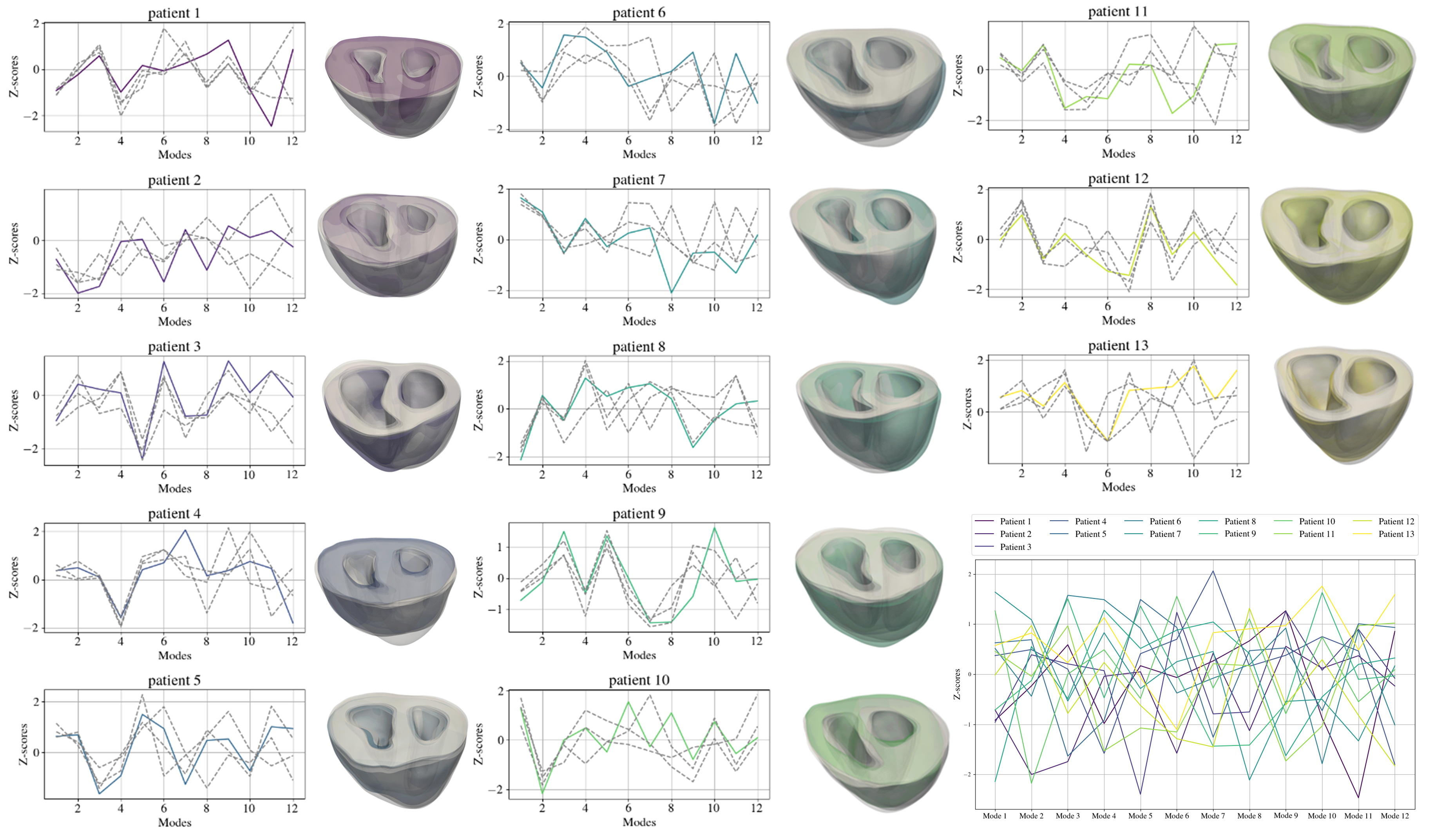}
    \caption{Parallel plots and overlay meshes comparing original and generated patients. Solid colored lines represent original cohort z-scores, while dashed lines indicate generated synthetic samples closest to each subject.}
    \label{fig:parallel_plots_synthetic_cohort}
\end{figure}

\medskip \noindent
Using Latin Hypercube Sampling, we generate an extensive pool of 10,000 candidate samples. 
After filtering these samples according to the estimated probability density function from the original cohort, we retain 294 realistic synthetic samples. 
Figure \ref{fig:parallel_plots_synthetic_cohort} visualizes these results using parallel plots.  The solid colored lines represent the ensemble of z-scores from the original patient cohort, while the dashed lines illustrate the selected synthetic samples closest to each patient. For further clarity, we also provide overlays between each original anatomy and its four nearest synthetic counterparts, highlighting the anatomical similarities and variations captured by the synthetic data generation method.

\subsection{Electrophysiology simulations}
From our electrophysiology simulations, we can measure the time required for complete depolarization across the entire biventricular geometry. A position (x,y,z) is considered depolarized when its action potential exceeds -10 mV. The time required for a node to reach depolarization is referred to as its activation time. Across cohorts, the shortest duration recorded for complete depolarization of the geometry was 64.5 ms (patient 19), while the longest was 92.7 ms (patient 7).

\begin{figure}[H]
  \centering
    \includegraphics[width=\textwidth]{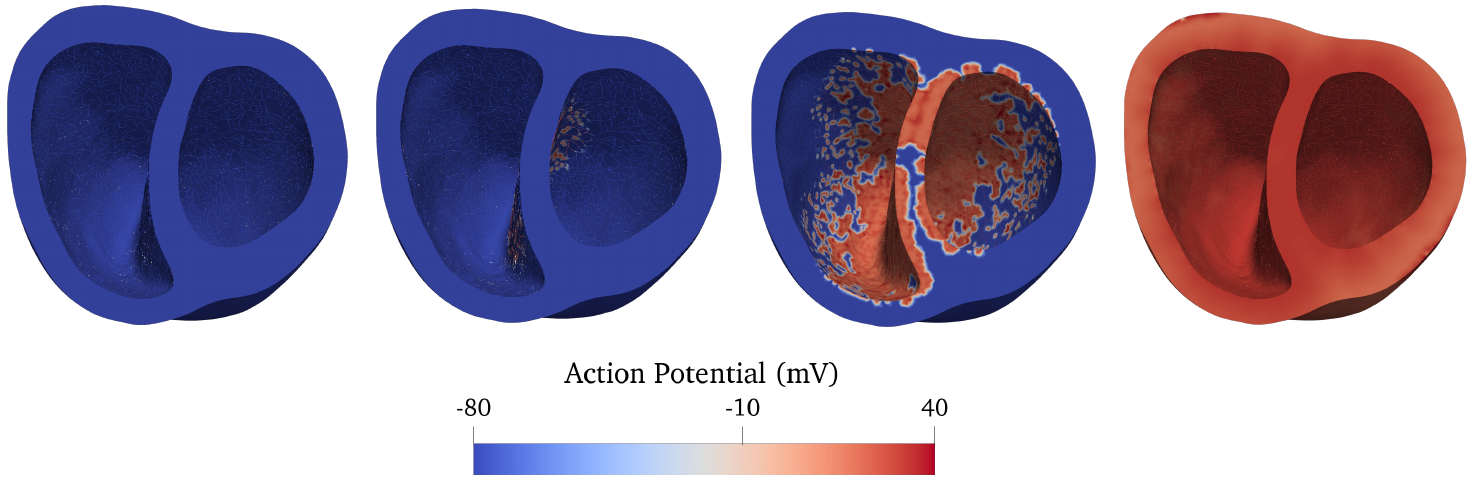}
     \caption{Action potential progression for synthetic patient 19 at sequential time points: $t= 0.1$ ms, $t = 16.1$ ms, $t = 32.3$ ms, $t = 64.5$ ms.}
   \label{action_potentials_19}
\end{figure}

\begin{figure}[H]
  \centering
    \includegraphics[width=\textwidth]{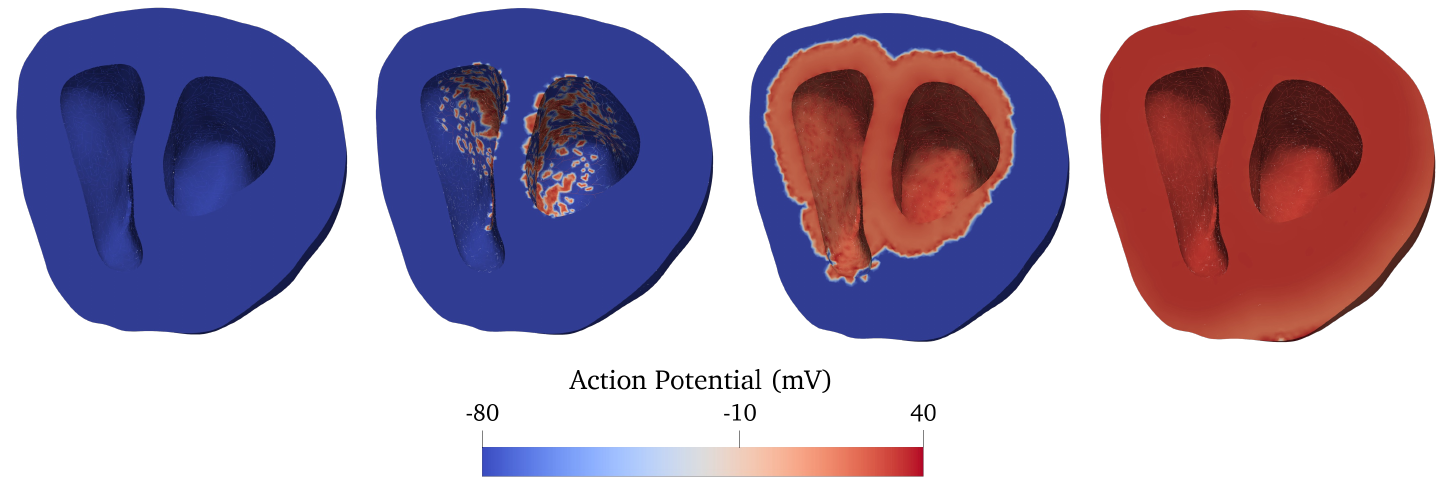}
   \caption{Action potential progression for synthetic patient 7 at sequential time points: $t= 0.1$ ms, $t = 23.2$ ms, $t = 46.4$ ms, $t = 92.7$ ms.}
   \label{action_potentials_7}
\end{figure}

Figures~\ref{action_potentials_19} and~\ref{action_potentials_7} illustrate the action potentials at different time steps for these patients. These figures depict the action potential at 0.1 ms, and approximately 1/4, 1/2, and the full duration required to achieve complete depolarization, highlighting the temporal progression of depolarization across the biventricular geometry. In the synthetic patient cohort, the shortest duration recorded for complete depolarization of the geometry was 64.5 ms, while the longest was 92.7 ms. In the original patient cohort, the durations for complete biventricular depolarization ranged from 67 ms to 88 ms.

This variation in depolarization times highlights the differences in electrophysiological behavior across patient geometries, with each geometry exhibiting a unique depolarization rate. To focus on the impact of anatomical differences, we intentionally used the same electrophysiological parameters across all geometries, ensuring that variations in depolarization dynamics could be attributed solely to structural factors. However, the resulting activation maps have not been clinically validated against direct patient measurements. The physics-based model used in this study serves as a reference for assessing activation patterns. 

\begin{figure}
  \centering
    \includegraphics[width=\textwidth]{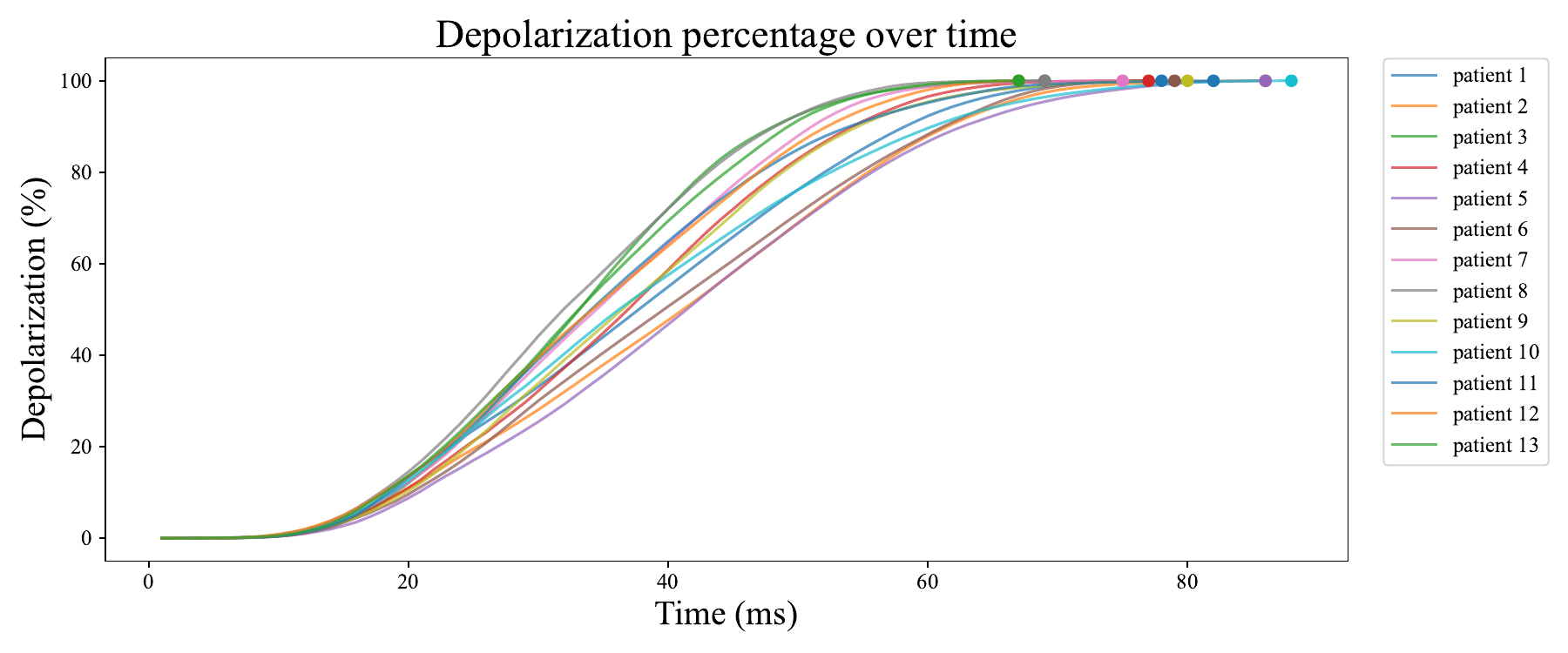}
   \caption{Percentage of depolarized points within the biventricular geometry of each patient over time measured in ms. Dots on each line mark the point at which 100$\%$ depolarization is reached.}
   \label{depolarization_percent}
\end{figure}

The impact of these geometric differences is further illustrated in Figure~\ref{depolarization_percent}, which charts the percentage of points that have been depolarized over time in the original patient cohort. Notably, complete depolarization times range from 67 ms to 88 ms, highlighting the variability in both the rate and timing of depolarization. Figure~\ref{depolarization_percent} also reveals that while some geometries depolarize at a nearly constant rate, others exhibit periods of rapid or delayed activation.  

\begin{figure}[H]
  \centering
    \includegraphics[width=\textwidth]{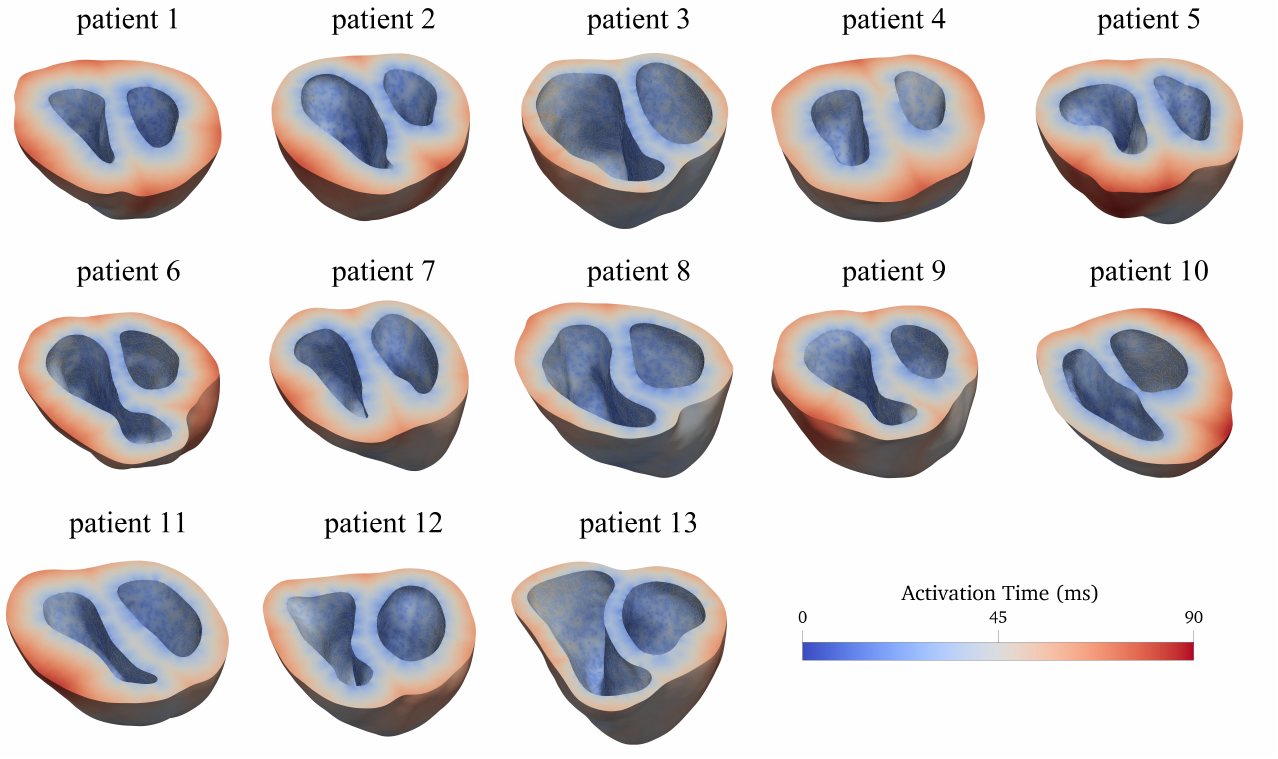}
   \caption{Simulated activation maps of the original cohort.}
   \label{activation_maps}
\end{figure}

Figure~\ref{activation_maps} presents the activation maps for the 13 original patient geometries, where each map visually encodes activation time using a color gradient --- blue represents early activation, while red indicates later activation. 
These visualizations enable us to identify regions of rapid activation (e.g. Purkinje fibers) and areas of delayed activation (e.g the ventricular myocardium). 

\begin{figure}[H]
  \centering
    \includegraphics[width=\textwidth]{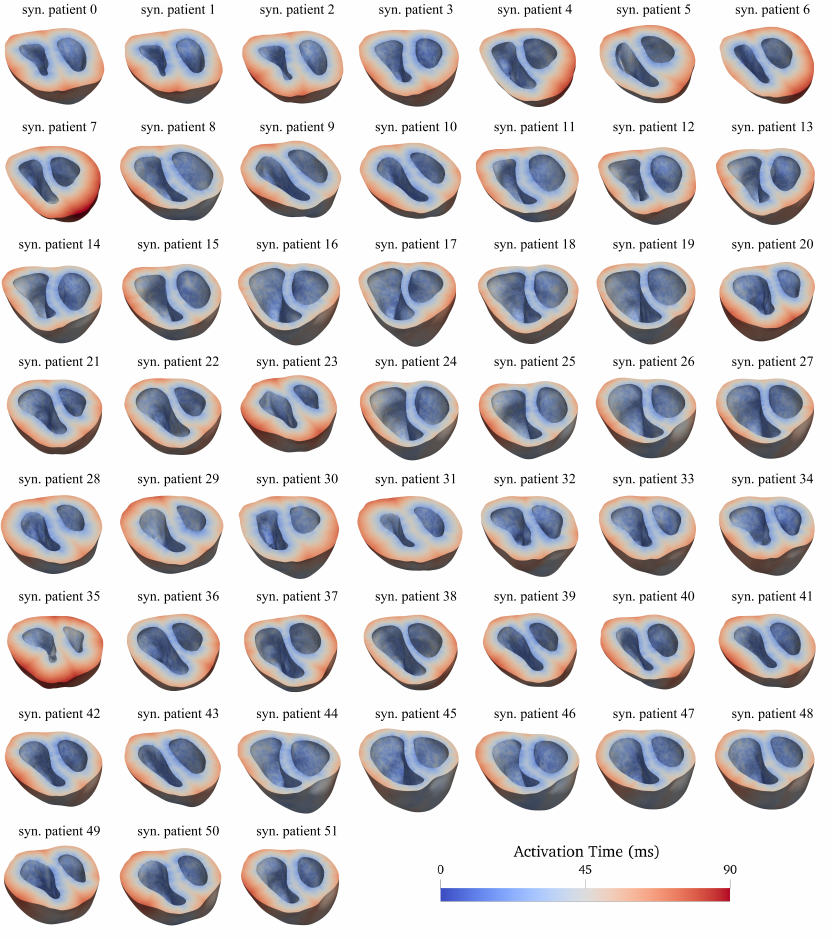}
   \caption{Simulated activation maps of the synthetic cohort.}
   \label{synthetic_activation_maps}
\end{figure}

Figure~\ref{synthetic_activation_maps} provides the corresponding activation maps for the synthetic cohort. Both figures highlight the variability in electrical signal propagation across the biventricular geometries. These variations emphasize the individualized nature of cardiac electrophysiology and the inherent heterogeneity present in both cohorts.

\subsection{Branched Latent Neural Maps Results}
We train the BLNM by using an optimal configuration of 16 layers, 42 neurons per layer, 5 states, a disentanglement level of 2, and a learning rate of 0.0005. For each epoch, we randomly select 1,485 points from our training geometries. The BLNM model runs for 1,000 epochs with the objective of predicting the activation time for all nodes within the geometry, specifically aiming to minimize MSE. 

Training our BLNM surrogate model on a single Apple M3 Pro (12-core) CPU took approximately 12 minutes (711 seconds) in total.  The average inference time per prediction is 0.245 ms, computed by running predictions on each geometry in our testing cohort.

We achieve the best overall training MSE of 0.0012 and the best overall testing MSE of 0.0034. The MSE across both training and testing cohorts maintain an order of magnitude of $10^{-3}$, with a couple of training geometries producing an MSE of magnitude $10^{-4}$. Each individual training geometry, from synthetic patient 1 to synthetic patient 52, records a loss ranging between 0.0009 and 0.0014. The range of losses for our testing cohort varies, with patient 12 demonstrating a loss of  0.0017, while patient 6 exhibits the highest testing loss of 0.0051. Table~\ref{label_error} provides a detailed error analysis for each geometry in the testing cohort.

\begin{table}[h]
    \centering
    \setlength{\tabcolsep}{12pt}
    \renewcommand{\arraystretch}{1.2}
    \begin{tabular}{| c | c | c | c |}
    \hline
    \textbf{Patient Label} & \textbf{Test Error} & 
    \textbf{Patient Label} & \textbf{Test Error} \\ 
    \hline
    patient 1 & 0.0021 & patient 2 & 0.0046 \\ \hline
    patient 3 & 0.0028 & patient 4 & 0.0049 \\ \hline
    patient 5 & 0.0050 & patient 6 & 0.0051 \\ \hline 
    patient 7 & 0.0045 & patient 8 & 0.0028 \\ \hline
    patient 9 & 0.0036 & patient 10 & 0.0031 \\ \hline patient 11 & 0.0031 & patient 12 & 0.0017 \\ \hline
    patient 13 & 0.0023 &  &\\ \hline
    \end{tabular}
    \captionsetup{font = small}
    \caption{BLNM MSE for each patient of the original cohort.}
    \label{label_error}
\end{table}

\begin{figure}[h]
  \centering
    \includegraphics[width=\textwidth]{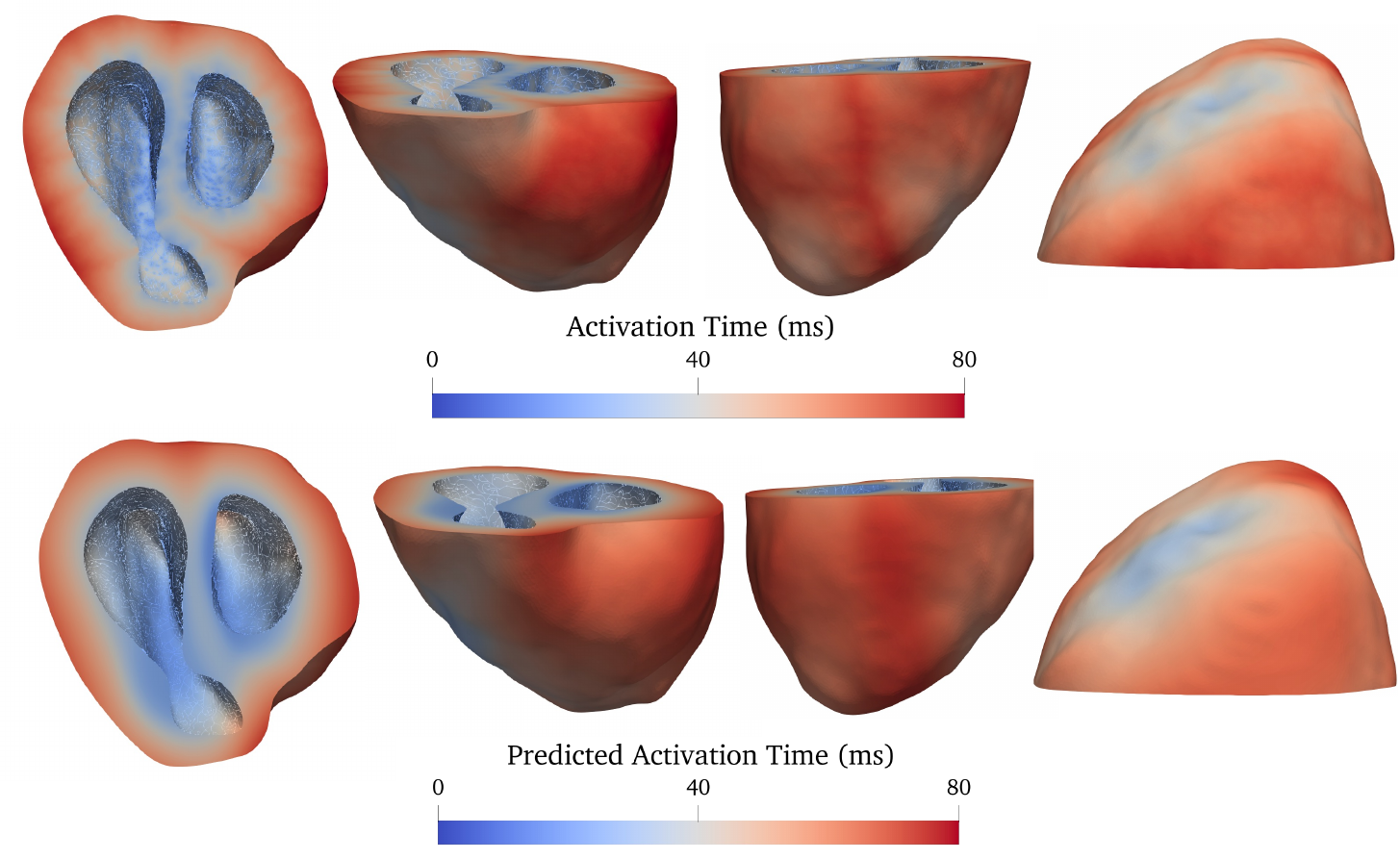}
   \caption{Simulated activation map (top row) vs. BLNM prediction (bottom row) for patient 6.}
   \label{activation_maps_1}
\end{figure}

\begin{figure}[H]
  \centering
    \includegraphics[width=\textwidth]{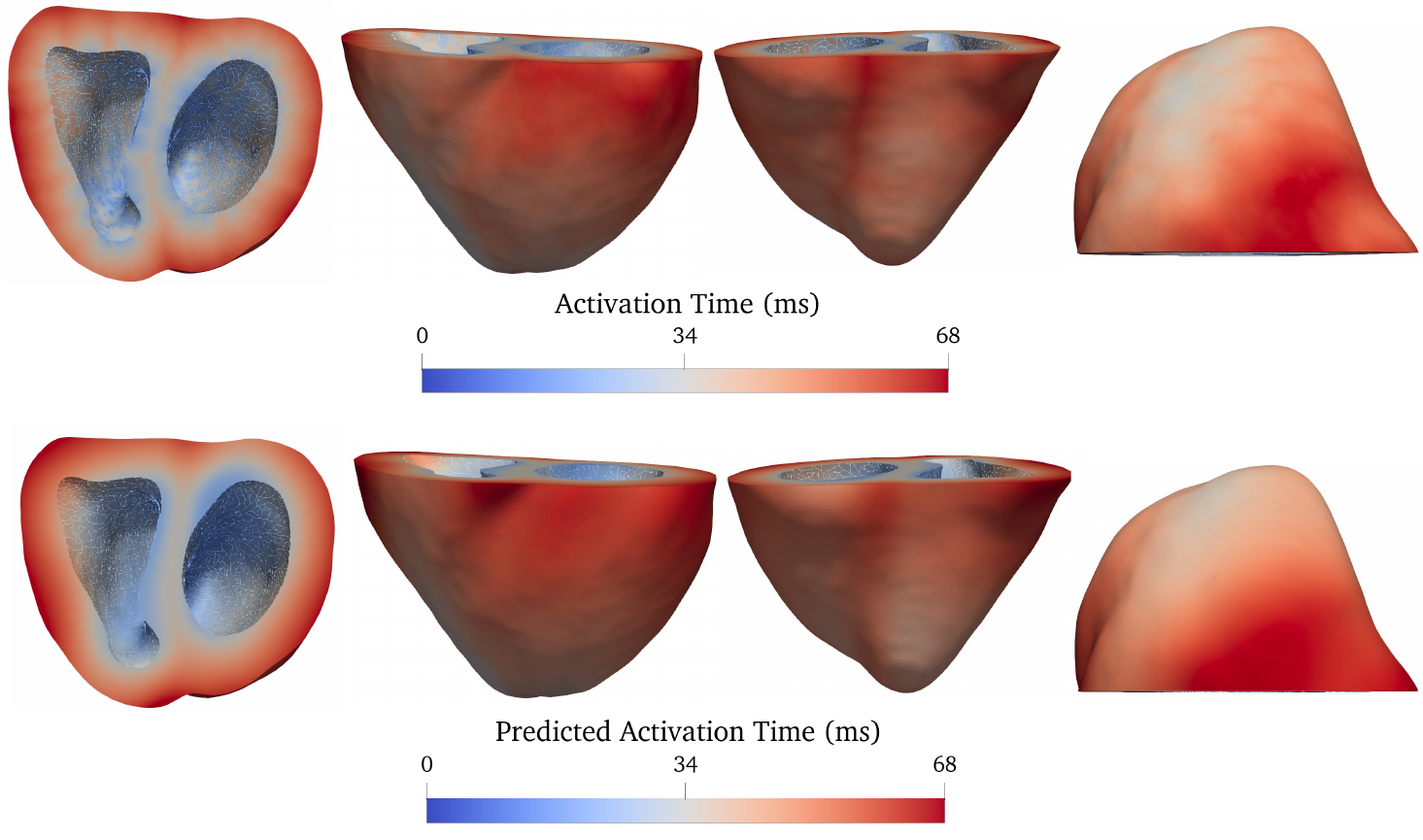}
   \caption{Simulated activation map (top row) vs. BLNM prediction (bottom row) for patient 12.}
   \label{activation_maps_2}
\end{figure}

Figures~\ref{activation_maps_1} and~\ref{activation_maps_2} illustrate the actual and predicted activation times for our worst-performing testing geometry (patient 6) and our best-performing testing geometry (patient 12), respectively. Each figure presents activation times from four distinct perspectives, offering a comprehensive view of the spatial distribution of electrophysiological activity in these 3D activation maps. In both images, the top row displays the activation times obtained from our electrophysiology simulation, while the bottom row shows the activation times predicted by our BLNM model. The color gradient represents the full range of activation times across the biventricular geometry, illustrating regional variations in conduction and the complex spatial distribution of electrophysiological activity.

This pictoral comparison reveals a closer alignment between the predicted and simulated activation times for patient 12, while patient 6 exhibits larger discrepancies, reflecting the differences captured by our MSE values. 
Nonetheless, in both geometries we see that the BLNM prediction exhibits a smoother gradient transition in comparison to the numerical simulation. This indicates that while the BLNM captures well most of the activation pattern, it tends to average out the finer details. Latent dynamics networks prioritize stable global representations over fine-grained local variations. 


We now focus on identifying the specific regions within each testing patient geometry that exhibit significant discrepancies between the simulated and predicted activation times. The prediction errors, defined as a difference in ms between the simulated and actual activation times, range from a maximum value of 18.8 ms to 27.4 ms across patients. 

\begin{figure}[H]
  \centering
    \includegraphics[width=\textwidth]{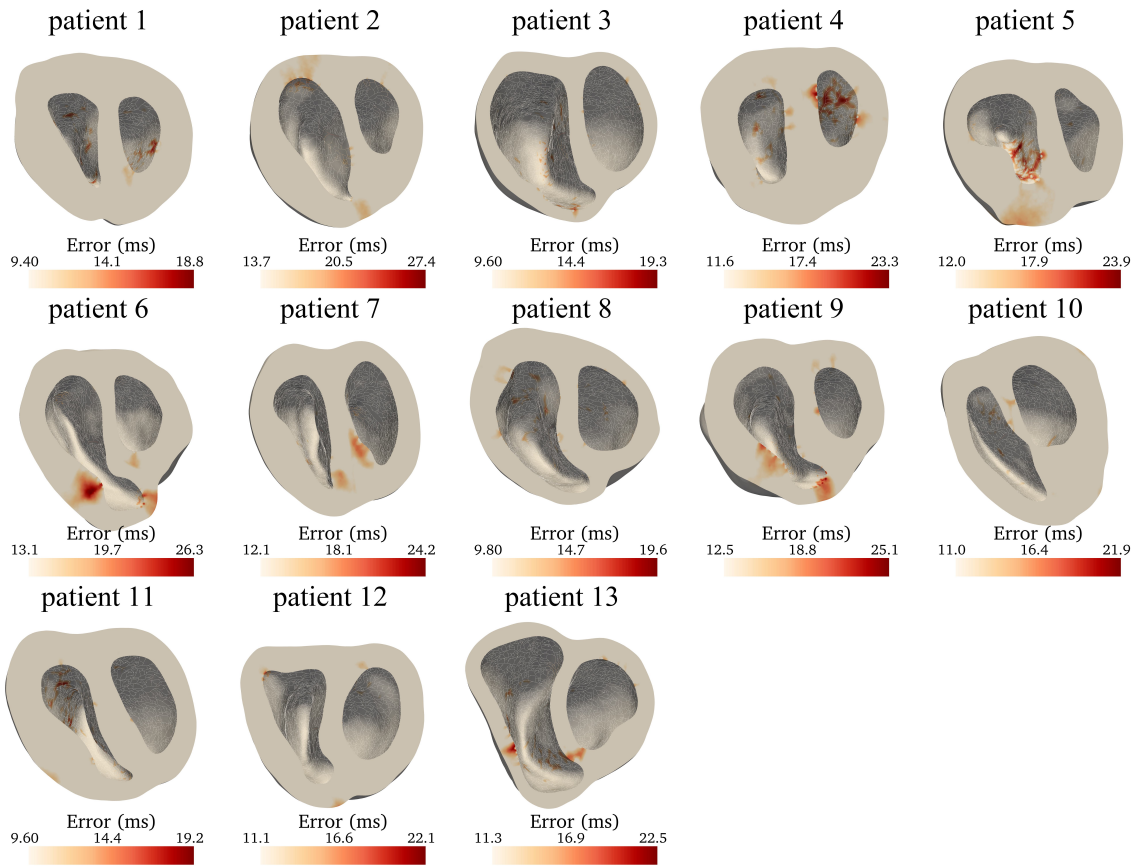}
   \caption{Top $50\%$ of absolute errors between predicted and simulated activation times for the original cohort. The color scale ranges from white, indicating areas where the prediction error is at or below the median, to red, where errors reach the maximum for that particular geometry.}
   \label{error_image}
\end{figure}

Figure~\ref{error_image}
illustrates the geometries in the original cohort, each colored-coded to highlight
the top $50\%$ of prediction errors. This visualization helps identify regions where the BLNM struggles most in accurately predicting cardiac activation times.
We observe that areas with higher errors often coincide with structurally complex regions of the heart. For example in patients 6,7, and 9, we observe higher errors in parts of the geometry where the right ventricle is particularly narrow. Additionally, each geometry demonstrates significant errors at the initial points of the Purkinje network. We also observe that our BLNM predicts significantly higher activation times for the nodes in the first branch --- the main segment immediately following the first Purkinje node before splitting into other branches. This occurs in both the LV and RV Purkinje networks.

\begin{figure}[H]
  \centering
    \includegraphics[width=\textwidth]{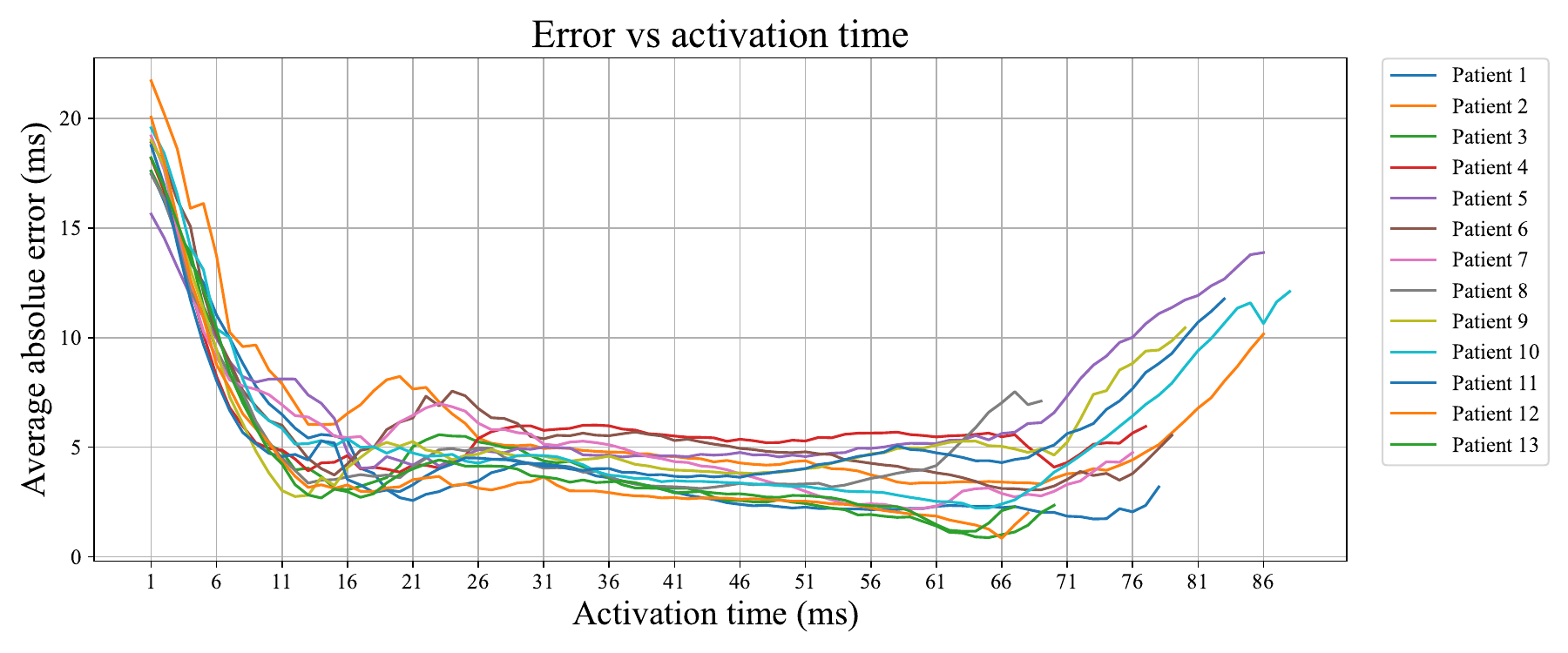}
   \caption{Average error for all nodes sharing an activation time within a geometry, with each line representing a different patient from patients 1 through 13.}
   \label{error_vs_activation}
\end{figure}

 Figure~\ref{error_vs_activation} further supports the conclusion that we obtain higher errors at the initial points of the Purkinje. In this figure we compare the average error for every node across a geometry that shares an activation time. For each of the 13 geometries, we notice significantly higher discrepancies between the simulated and predicted activation times at nodes with low activation times. Action potential is applied at time $t = 0$ to the nodes that initiate our Purkinje networks. These points have the lowest activation times and, as evidenced by the plot, exhibit the highest error. Error generally decreases as activation time increases but rises again for the last activating nodes, though still exhibiting smaller absolute errors than the initial activation times.

The geometrical variability of our cohort can be effectively explained using shape scores $\alpha_{i,m}$,
which describe the contribution of each principal mode of variation to a given subject, $S_i$. These shape coefficients are normalized as z-scores. Investigating how the number of modes used in the surrogate model impacts its performance is essential for understanding the extent of variability that can be captured and determining the number of z-scores needed to effectively represent the data. We evaluate model accuracy while systematically varying the number of modes used in our surrogate model and maintaining the same optimal hyperparameters previously determined. 

From Table 4 we see that adding more modes generally reduces the error between predicted and simulated activation time, indicating improved model performance. However, beyond around 5 modes, the rate of improvement diminishes, and additional modes capture less meaningful variations. Nonetheless, the best losses across all testing geometries are ultimately achieved using all 12 modes. Through this convergence study, we evaluate how well our surrogate model encodes and generalizes the geometrical variability within the testing cohort.

\begin{table}[H]
    \centering
    \setlength{\tabcolsep}{9pt}
    \renewcommand{\arraystretch}{1.2}
    \begin{tabular}{| c | c | c |c | c|}
    \hline
    \textbf{$\#$ modes} & \textbf{Training Loss} & 
    \textbf{Testing Loss} & \textbf{Worst Geom Loss} & \textbf{Best Geom Loss }\\ 
    \hline
    1 & 0.0056 & 0.0079 & 0.0154 & 0.0047  \\ \hline
    2 & 0.0043 & 0.0062 & 0.0112 & 0.0042\\ \hline
    3 & 0.0030 & 0.0053 & 0.0095 &  0.0038 \\ \hline
    4 & 0.0022 & 0.0048 & 0.0091 & 0.0035 \\ \hline
    5 & 0.0016 & 0.0039 & 0.0051 & 0.0023 \\ \hline
    6 & 0.0015 & 0.0038 & 0.0063 & 0.0023\\ \hline
    7 & 0.0012 & 0.0038 & 0.0051 & 0.0022  \\ \hline
    8 & 0.0012 & 0.0036 & 0.0053 & 0.0022  \\ \hline
    9 & 0.0012 & 0.0036 & 0.0066 & 0.0020 \\ \hline
    10 & 0.0013 & 0.0038 & 0.0061 & 0.0020 \\ \hline
    11 & 0.0014 & 0.0039 & 0.0079 & 0.0024 \\ \hline
    12 & 0.0012 & 0.0035 & 0.0051 & 0.0017 \\ \hline
    \end{tabular}
    \captionsetup{font = small}
    \caption{MSE for the BLNM surrogate model across varying numbers of modes. The table reports the overall best training and testing loss, as well as the testing loss of the worst-performing and best-performing geometries, when utilizing between 1 and 12 modes of variation.}
    \label{label_error}
\end{table}

	\section{Discussion}
\label{sec:discussion}
Accurate modeling of cardiac anatomy and function is essential for advancing the understanding and treatment of congenital heart disease. Neural networks, in particular, have emerged as powerful tools for capturing spatial dynamics and approximating key aspects of cardiac hemodynamics, including pressure, flow rates, cardiac mechanics, and electrodynamics \cite{Cicci2023, Regazzoni_3, Costabal2016, Peirlinck2019, SahliCostabal2019, Pegolotti2024, xie2022, xie2023}. By integrating physics-based and data-driven methods, our surrogate model enhanced computational efficiency while preserving prediction accuracy, facilitating progress towards clinically applicable, patient-specific models for understanding and treating congenital heart disease. 

Branched Latent Neural Maps (BLNMs) are a novel neural network architecture designed to model the spatial dynamics of physical processes by separating intrinsically different input spaces while leveraging latent variables to capture the underlying variability. These models enable the prediction of both spatial and temporal quantities by integrating diverse data modalities, including spatial coordinates, temporal dynamics, and structural and functional patient-specific characteristics. However, their application thus far has been limited to a single patient geometry for parameter calibration and in silico electrocardiogram generation \cite{Salvador2024BLNM}. To date, BLNMs have not been explored for cohort analysis or for predicting spatiotemporal dynamics with geometry-specific variables.

In this study, we extended the application of BLNMs to predict activation time using spatial coordinates and z-scores. To evaluate their ability to generalize across complex geometries, we utilized a cohort of ToF patients. Given its significant anatomical variability, ToF served as an ideal test case to assess the ability of the model to learn across diverse cardiac structures. We generated 13 biventricular meshes based on CT scans of pediatric ToF patients. These meshes exhibited significant differences in ventricular size, myocardial volume, wall thickness, and septal morphology --- factors that strongly influence cardiac  electrophysiology and function \cite{trayanova2024, chan2014}. The diversity and complexity of cardiac geometries in CHD patients are key reasons why current surrogate models for predicting cardiac function require retraining to generalize across patient-specific anatomies.  

Nonetheless, for our BLNM surrogate to perform effectively, a larger dataset than the original cohort is beneficial. To address this limitation, statistical shape modeling played a crucial role in augmenting our cohort of original patients. We leveraged an LDDMM-based approach to conduct anatomical mapping of the biventricular geometries in the original population, and we proposed a novel sampling approach to augment the original population whilst retaining the geometrical variability of the original cohort. 
Several strategies have been proposed for the generation of in-silico cohorts of patients \cite{verstraeten_2023_generation_aortic,kong2024}. In this regard, data-driven generation methods, including bootstrapping and Gaussian sampling, are particularly suited when the goal of the sampling is to reproduce the existing population \cite{romero_2021_synthetic_gen}. However, such methods perform poorly when the original population is small, and can result in unfeasible generated geometries \cite{romero_2021_synthetic_gen}. Acceptance criteria based on clinical biomarkers are often used to filter out improbable shapes, but are very challenging to compute especially for complex anatomies such as congenital heart disease patients. In this work, we proposed a novel generation strategy that allowed us to constrain the sample selection to the regions of the shape parameter space that are more representative of the underlying anatomical variability in the population. By selecting the four closest samples of this restricted distribution, we effectively augmented the original cohort to improve the performance of our surrogate model. Using this synthetic cohort for training and the original cohort for testing, we were able to overcome the challenge of working with a small number of patient geometries.

We conducted 65 electrophysiology simulations, generating activation maps for both the 13 original and 52 synthetic geometries. One key challenge was accurately representing the Purkinje network, as its intricate structure is difficult to image, making its inclusion in electrophysiology models particularly challenging \cite{Krishnamoorthi2014}. To address this, we refined an existing Purkinje algorithm by incorporating a standardized coverage requirement, ensuring that the network spans a consistent proportion of each and every ventricle. Additionally, we introduced a systematic approach for selecting the initial simulation point and determining the direction of the first branch in each network. This refinement is crucial, as the Purkinje network plays a fundamental role in the depolarization dynamics and the overall electrophysiological behavior. Moreover, by preserving randomized branching, we still captured the natural variability in human heart physiology.

Our resulting electrophysiology simulations revealed significant variability in activation times and depolarization dynamics in a diverse cohort of biventricular geometries. Notably, each patient geometry exhibited distinct depolarization timelines and rates. For example, patient 3 reached complete depolarization in 67 ms, while patient 10 required 87 ms. Our simulation parameters, including anisotropy, conductivities, and the ten Tusscher-Panfilow model, maintained the same across all geometries. Despite this consistency, metric analysis revealed that both patients fell within the midrange for ventricular surface area, volume, wall thickness, and apicobasal height. This suggests that depolarization dynamics are influenced by a combination of geometric factors rather than one dominant anatomical parameter. In this study, we intentionally focused on isolating the effects of geometry on depolarization dynamics by maintaining uniform electrophysiological properties across all cases. However, we acknowledge that additional sources of variability such as fibrosis, tissue heterogeneity, or differences in ion channel distributions can impact conduction properties \cite{verheule2021}. The inclusion of these factors represents an avenue for future research that could provide a more comprehensive understanding of activation variability.

We reported the results of the BLNM surrogate model using hyperparameters determined using OptunaSearch, a hyperparameter optimization algorithm within Ray Tune. Our results demonstrated the ability of BLNMs to predict activation times across a range of complex biventricular geometries with high accuracy relative to physics-based models, which are used as ground truth for evaluation, while also maintaining computational efficiency. 

A visual comparison between the simulated and BLNM-predicted activation maps in Figures~\ref{activation_maps_1} and \ref{activation_maps_2} showed that our model effectively captured the general spatial distribution of activation times. However, the predicted results exhibited smoother gradients compared to the simulated results, where activation follows a more localized variation with sharper gradients. This smoothing effect causes activation values in the endocardium, particularly near Purkinje nodes, to be influenced by a gradient diffusion effect, reducing contrast. To mitigate this limitation, Fourier embeddings can be incorporated for input space coordinates \cite{tancik2020}. By leveraging Fourier feature mappings, the model can better represent sharp transitions and localized activation variations, minimizing the averaging effect. These embeddings can enhance the model's ability to capture high-frequency functions rather than favoring smooth features during training. However, this improvement comes at the cost of an increased computational complexity \cite{ldnets}.

Furthermore, by analyzing prediction errors, measured as the absolute difference between predicted and simulated activation times (in ms), we identified the regions where BLNMs struggled the most to accurately predict activation times. One possible explanation for the largest errors occurring at the earliest activation times, specifically at the first branch of the Purkinje network, is that the algorithm required varying initial branch lengths to ensure adequate coverage of the network. While these early activation regions exhibit the highest discrepancies, small inaccuracies in activation can propagate through the network, affecting downstream activation times. This highlights the importance of accurately modeling the initial activation points and propagation patterns of the Purkinje network to minimize cumulative errors.

Additionally, while prediction errors decrease and remain consistent after the initial Purkinje nodes, we observed a secondary spike in prediction error toward the end of activation. This is visually supported by high errors at the base of the geometries, particularly in regions with greater structural complexity, such as a narrow RV. This issue arises due to the 1D-3D coupling problem, where our fractal tree-based Purkinje network is projected onto a 3D surface, making it difficult for the BLNM to accurately capture activation propagation near the base of the geometry. Namely, the BLNM struggled to reproduce the abrupt termination of the fast conducting Purkinje network. These errors are not solely a limitation of the BLNM, but also highlight the critical role the Purkinje network plays in cardiac activation \cite{Alvarez2025}. Despite this, even though the Purkinje network is generated independently for each geometry, BLNMs successfully capture the activation maps of the 1D-3D Purkinje-myocardium system with MSE of order $10^{-3}$. Incorporating more detailed modeling of these networks, such as patient-specific variability in branch lengths and angles could significantly improve predictions. Given the complexity of Purkinje activation, the ability of BLNMs to generalize efficiently while maintaining computational efficiency is a significant achievement. Although challenges remain to accurately capture activation times in these complex regions, our results highlight the robustness and adaptability of the model, with potential for further refinement.

To further assess model performance, we varied the number of anatomical variability modes used in our surrogate model while maintaining the same BLNM architecture. The results, summarized in Table~\ref{label_error}, indicate that increasing the number of anatomical modes generally enhances model performance, reducing both training and testing errors. With only a few modes, the model primarily captures large-scale or global variations, such as overall size or orientation, but does not account for finer structural details. As more modes are incorporated, the model becomes capable of encoding more nuanced and localized variations, such as curvature differences or small regional deformations. Using all 12 modes, which performed best across all metrics, allows the model to generalize effectively to unseen shapes by capturing the full spectrum of anatomical variability. In general, our analysis demonstrates the critical role of the number of modes in determining the performance of the surrogate model, with the 12 modes providing the most complete representation.
    \section{Conclusions}
\label{sec:conclusions}

In this study, we evaluated the performance of a Branched Latent Neural Maps (BLNM) surrogate model in predicting activation times generated by physics-based models across a diverse and complex cohort of biventricular geometries derived from patients with Tetralogy of Fallot (ToF). By leveraging both patient-derived and synthetic datasets, our analysis highlights the importance of capturing anatomical variability to develop a robust and generalizable surrogate model. We found that BLNMs generate activation times given space coordinates and z-scores as inputs with high accuracy and computational efficiency, achieving low MSE values of order $10^{-3}$ across all geometries.  

The innovations presented in this study span both methodology and application. The inclusion of synthetic geometries generated through statistical shape modeling (SSM) enabled us to address the limitations of working with a small and challenging cohort, thereby improving the model's ability to generalize across highly varying geometries. Furthermore, we applied BLNMs to predict geometry-specific activation times, demonstrating that we can develop a surrogate model for electrophysiology that does not require retraining across different geometries. By optimizing hyperparameters, such as network depth, neurons per layer, and disentanglement levels we achieved an MSE of 0.0034, with errors ranging from 0.0017 to 0.0051 across testing geometries. The training process, which completed in approximately 12 minutes (711 seconds) on a single CPU, highlights the computational efficiency of our approach. This rapid training time, combined with the ability of the model to maintain a low MSE across both training and testing cohorts, underscores the potential of BLNMs for large-scale applications, including cohort studies and personalized cardiac simulations.

Despite these achievements, our analysis identified key areas for improvement, particularly in capturing the Purkinje network variability and refining activation time  prediction in regions of high structural complexity. The variability in prediction errors further highlights the sensitivity of cardiac activation dynamics to geometric features, such as myocardial wall thickness, septal morphology, and ventricular size. To enhance the model's predictive capabilities future work should focus on improving patient-specific Purkinje network representation and refining the model's ability to capture sharp transitions in activation by incorporating Fourier embeddings. 

Additionally, while our results demonstrate that the surrogate model can accurately predict activation patterns within the framework of our physics-based simulations, clinical accuracy depends on the simulations reliably representing electrophysiological behavior. Therefore, verifying the simulated activation patterns against clinical activation maps is essential to ensure their accuracy. Comparing the BLNM predictions to patient ECG signals, particularly in individuals with CHD, would help evaluate the clinical relevance of the model and its ability to capture key electrophysiological features. 
Nevertheless, our findings establish that the model can successfully learn activation dynamics, reinforcing its potential for patient-specific cardiac modeling. 

In conclusion, this study advances the application of BLNMs for interpatient analysis, demonstrating their potential for accurately predicting cardiac activation times across anatomically diverse biventricular geometries. Future work should prioritize integrating more detailed patient-specific anatomical data, particularly regarding the Purkinje network, and exploring larger, more diverse datasets to enhance the robustness and clinical applicability of surrogate modeling in cardiac electrophysiology. Additionally, we propose incorporating multiple pathological conditions, such as hypoplastic left heart syndrome (HLHS) and transposition of the great arteries (TGA), into the same network. This would enable the model to generalize across multiple independent patient cohorts, making it a more versatile tool for electrophysiological modeling and potential clinical applications.

	\section*{Acknowledgements}

This work was supported by the National Science Foundation (DGE-2146755 to E.M and 2310909 to M.S, F.K, A.L.M); the National Institutes of Health (Grant No. R01HL173845 to M.S, F.K, A.L.M); the Research Foundation – Flanders, Fonds voor Wetenschappelijk Onderzoek –Vlaanderen (Grant No. 11PS524N, to B.M.) and European Union’s Horizon Europe Research and Innovation Program (VITAL - Grant No. 101136728, to M.P.). We acknowledge the Stanford Cardiovascular Institute.

    \newpage
    \printbibliography

\end{document}